\documentclass[3p, review]{elsarticle}  

\usepackage{microtype}
\usepackage{xcolor}

\usepackage{amsfonts}
\usepackage{amsmath}
\usepackage{amssymb}
\usepackage{pifont}
\usepackage{lipsum}
\usepackage{mathrsfs}
\usepackage{cancel}
\usepackage{bbm}
\usepackage{siunitx}
\usepackage{mathtools}

\usepackage{graphicx}
\usepackage{caption,subcaption}

\newcommand\numberthis{\addtocounter{equation}{1}\tag{\theequation}}

\usepackage{lineno}
\modulolinenumbers[2]

\usepackage{hyperref}

\newcommand*\patchAmsMathEnvironmentForLineno[1]{%
  \expandafter\let\csname old#1\expandafter\endcsname\csname #1\endcsname
  \expandafter\let\csname oldend#1\expandafter\endcsname\csname end#1\endcsname
  \renewenvironment{#1}%
     {\linenomath\csname old#1\endcsname}%
     {\csname oldend#1\endcsname\endlinenomath}}%
\newcommand*\patchBothAmsMathEnvironmentsForLineno[1]{%
  \patchAmsMathEnvironmentForLineno{#1}%
  \patchAmsMathEnvironmentForLineno{#1*}}%
\AtBeginDocument{%
\patchBothAmsMathEnvironmentsForLineno{equation}%
\patchBothAmsMathEnvironmentsForLineno{align}%
\patchBothAmsMathEnvironmentsForLineno{flalign}%
\patchBothAmsMathEnvironmentsForLineno{alignat}%
\patchBothAmsMathEnvironmentsForLineno{gather}%
\patchBothAmsMathEnvironmentsForLineno{multline}%
}

\journal{Signal Processing}

\begin{document}

\begin{frontmatter}

\title{Efficient approximations of the multi-sensor labelled multi-Bernoulli filter}

\author[su]{S.C.J. Robertson}
\ead{16579852@sun.ac.za}
\author[su]{C.E. van Daalen}
\ead{cvdaalen@sun.ac.za}
\author[su]{J.A. du Preez}
\ead{dupreez@sun.ac.za}

\address[su]{Department of Electrical and Electronic Engineering, Stellenbosch University, South Africa}

\begin{abstract}
In this paper, we propose two efficient, approximate formulations of the multi-sensor labelled multi-Bernoulli (LMB) filter, which both allow the sensors' measurement updates to be computed in parallel.
Our first filter is based on the direct mathematical manipulation of the multi-sensor, multi-object Bayes filter's posterior distribution.
Unfortunately, it requires the division of probability distributions and its extension beyond linear Gaussian applications is not obvious.
Our second filter approximates the multi-sensor, multi-object Bayes filter's posterior distribution using the geometric average of each sensor's measurement-updated distribution.
This filter can be used under non-linear conditions; however, it is not as accurate as our first filter.
In both cases, we approximate the LMB filter's measurement update using an existing loopy belief propagation algorithm.
Both filters have a constant complexity in the number of sensors, and linear complexity in both number of measurements and objects.
This is an improvement on an iterated-corrector LMB (IC-LMB) filter, which has linear complexity in the number of sensors.
The proposed filters are of interest when tracking many objects using several sensors, where filter run-time is more important than filter accuracy.
Simulations indicate that the filters' loss of accuracy compared to the IC-LMB filter is not significant.
\end{abstract}


\end{frontmatter}


\section{Introduction}
Multi-object tracking (MOT) is the process of jointly estimating an unknown, time-varying number of objects and their kinematic states using noisy sensor data \cite{blackman1999design}.
MOT is typically viewed from a probabilistic perspective, and the number of objects, their kinematic state and identity are inferred from the data.
The use of multiple sensors is common, as more data greatly reduces the uncertainty about both an object's existence and its state.
The general multi-sensor MOT formulation is computationally intractable, and drastic, but principled, approximation is required to produce practical solutions.

MOT has numerous applications and it is an established field of research \cite{blackman1999design, bar1975tracking, fortmann1983sonar}.
There are three main approaches in the literature: Multiple hypothesis tracking \cite{blackman2004multiple}, joint probabilistic data association \cite{fortmann1983sonar} and random finite sets (RFSs) \cite{mahler_book_2007, mahler_book_2014}.
Pioneered by Mahler, the RFS framework provides a unified approach to MOT and admits mathematically consistent, Bayesian solutions.
The theoretical underpinning of the framework is the multi-object Bayes filter, a direct generalisation of the single-object Bayes filter\cite{mahler_book_2007, mahler_book_2014}.
It has led to the development of tractable, approximate solutions such as the probability hypothesis density (PHD) filter \cite{mahler_phd_2003}, the cardinalised PHD filter \cite{mahler2007phd} and the cardinality-balanced multi-object multi-Bernoulli filter \cite{vo_vo_2009_cbmember}.

The aforementioned RFS-based filters are not true multi-object trackers, as they do not account for object identity and cannot produce a unique, coherent trajectory for any one object.
To remedy this, Vo and Vo introduced labelled RFSs, which allows object identity to be addressed in a mathematically consistent manner \cite{vo_vo_lrfs_2013}, unlike previous heuristic approaches.
Each object is uniquely associated with a discrete label which allows the objects' trajectories to be distinguished from one another.
This approach led to the development of the generalised labelled multi-Bernoulli (GLMB) filter, an exact, closed-form solution to the multi-object Bayes filter \cite{vo_vo_lrfs_2013, vo2014labeled}.
The GLMB filter is computationally demanding, as the number of hypotheses it holds over the possible number of objects and their states increases exponentially with time.
Data association is an NP-hard ranked assignment problem.
The labelled multi-Bernoulli (LMB) filter was developed as a computationally cheaper, approximate alternative \cite{vo_vo_lmb_2014}.
Unfortunately, the LMB filter still requires the computation of a GLMB posterior distribution before approximating it, thus further approximations are still required to make the filter efficient.

The GLMB filter has been extended to multiple sensors, but its implementation also requires the solution of an NP-hard ranked assignment problem \cite{vo_multisensor_2017}.
Similar to the single-sensor GLMB filter, a multi-sensor LMB filter can be derived from the multi-sensor GLMB filter and used as a computationally cheaper alternative.
However, it still faces the same computational hurdle as the multi-sensor GLMB filter.
Therefore, finding computationally efficient approximations and implementations of the multi-sensor LMB filter is both necessary and important.
In this paper, we propose two computationally efficient, parallelisable formulations of the multi-sensor LMB filter. 

\subsection{Survey of related work}
In this section, we briefly survey existing, efficient formulations of the single-sensor GLMB and LMB filters, before reviewing their respective multi-sensor formulations.
The single-sensor GLMB filter has been given an implementation that combines both its prediction and measurement update steps into a single step, and uses a Gibbs sampler to more efficiently approximate the NP-hard problem of data association \cite{vo2016efficient}.
In contrast to earlier implementations, this approach has a linear complexity in the number of measurements, and a quadratic complexity in the number of objects.
The single-sensor LMB filter can also be implemented using the same approach, with the same computational complexity \cite{Reuter_gibbs_2017}.
To further improve both the GLMB and LMB filters' efficiency, loopy belief propagation (LBP) has been used to resolve the data association problem with a lower computational complexity \cite{yang_wang_lbp_2018, meyer_2019_lmb_lbp}.
Developed by Williams et al., this LBP algorithm estimates each object's marginal association probability using message passing on a graphical model that describes the joint data association events, without the need to explicitly enumerate them \cite{williams_lau_lbp_2010}.
Remarkably, the algorithm reduces to fixed-point iteration, which is guaranteed to converge, and it is linear in both the number of objects and measurements \cite{williams_lau_con_2010}.
In this paper, we use Williams et al.'s algorithm to efficiently approximate the data association problem, but we define this message passing algorithm on a cluster graph, rather than a factor graph.
Graphical models have also been successfully applied to non-RFS-based multi-sensor multi-object tracking problems \cite{chen2003multitarget, chen2005distributed, chen2006data}.

The typical approach to multi-sensor, multi-object filtering is to perform each sensor's measurement update in succession \cite{vo_multisensor_2017}.
If no intermediate approximations are applied, then this approach is exact.
However, if approximations are applied between successive measurement updates, which is almost always practically necessary, then the filter's state estimates become dependent on the update order.
The multi-sensor GLMB filter has been given a sensor-order-independent implementation that also uses a joint prediction and measurement update step, and Gibbs sampling to resolve the data association \cite{vo_multisensor_2017}.
This implementation is linear in the number of sensors and quadratic in the number of objects, as is any LMB approximation of this multi-sensor GLMB filter.
To improve upon this complexity, we noted that if each sensor's measurement-updated distribution is approximated using an LMB distribution, then every sensor's measurement update can be computed independently and in parallel, before they are merged together to approximate the posterior distribution as an LMB distribution.
This can be achieved in one of two ways:
\begin{enumerate}
\item The mathematical manipulation of the multi-sensor, multi-object Bayes filter's posterior distribution.
This led us to the development of the parallel update LMB (PU-LMB) filter.
\item Approximating the multi-sensor, multi-object Bayes filter's posterior distribution as the weighted geometric average of each sensor's respective measurement-updated distribution.
This resulted in the development of the geometric average LMB (GA-LMB) filter.
\end{enumerate}
Both of these approaches result in a multi-sensor LMB filter with linear complexity in object and measurement number, and a constant complexity in number of sensors.
As the measurement updates are computed in parallel, neither filter depends on the order of the measurement updates. 

The first of our aforementioned approaches is based on the product multi-sensor PHD (PM-PHD) filter \cite{mahler_parallel_phd_2010}, which approximates each sensor's measurement-updated distribution using a multi-object Poisson distribution.
The PM-PHD approximation allows the multi-sensor, multi-object Bayes filter's posterior distribution to be approximated as Poisson.
Unlike the PM-PHD filter, the PU-LMB filter is capable of producing identifiable trajectories for each object.

The second approach is based on geometric average fusion, which is an established distributed multi-object, multi-sensor fusion algorithm \cite{li2019second}.
GA fusion approximates the posterior distribution using the weighted geometric average of each sensor's measurement-updated distribution, and it has been applied to the multi-sensor PHD filter \cite{battistelli2013consensus}.
The application of GA fusion to the LMB filter formulation produces unique, identifiable trajectories for each object, unlike the PHD filter.
GA fusion also coincides with generalised covariance intersection, which is a multi-object fusion rule first proposed by Mahler that extends covariance intersection and also involves the weighted geometric average of the measurement-updated distributions \cite{clark_fusion_2013}.
Generalised covariance intersection selects the optimal weights to be used in the geometric average \cite{uney2019fusion}, whereas GA fusion does not.

\subsection{Main contributions}
The contributions of this paper are:
\begin{enumerate}
\item A new derivation of Williams et al.'s LBP data association algorithm using a cluster graph, which we believe allows for a more concise encoding of the data association problem.
\item The theoretical development and implementation of a parallel update multi-sensor LMB (PU-LMB) filter, which approximates the multi-sensor, multi-object Bayes filter.
The implementation of the PU-LMB filter uses the LBP data association algorithm to approximate measurement-updated densities as LMB distributions.
Since the LPB data association algorithm does not enumerate association hypotheses, the PU-LMB filter has a low computational complexity.
\item The theoretical development and implementation of the geometric average LMB (GA-LMB) filter.
The filter is also approximated using the LBP data association algorithm, providing it with a low computational complexity.
\end{enumerate} 

This paper is organised as follows:
We first define labelled RFSs, before introducing the multi-sensor multi-object Bayes filter and approximating it using the LMB filter (Section \ref{section:foundations}).
We then review Williams et al.'s LBP data association algorithm using a new cluster graph derivation (Section \ref{section:efficient_approximation_lbp_lmb}).
With this foundational material in place, we then derive both the PU- and GA-LMB filters (Sections \ref{section:pu_lmb} and \ref{section:ci_lmb} respectively) and implement them (Section \ref{section:simulation_results}).
We compare our proposed filters to an iterated-corrector, multi-sensor LMB (IC-LMB) filter, and an LMB filter based on Vo et al.'s Gibbs sampler implementation of the multi-sensor GLMB filter \cite{vo_multisensor_2017}.

\section{Foundations}
\label{section:foundations}
In this section, we briefly review labelled random finite sets (RFSs) and the multi-sensor multi-object Bayes filter, with specific focus given to its approximation, the LMB filter.
For further details, we refer the reader to the original works \cite{vo_vo_lrfs_2013, vo_vo_lmb_2014}.
We directly build on this section's results when deriving the PU- and GA-LMB filters in Sections \ref{section:pu_lmb} and \ref{section:ci_lmb} respectively. 

In this paper, we denote an object's kinematic state -- its position, velocity, acceleration -- using lower-case letters ($x$).
An object's state takes on values in a state space $\mathbb{X}$, and we use blackboard bold letters to denote any space.
A multi-object state must describe both the number of objects and their respective states using a single variable; it is naturally represented by the set $X$ of single object states and we use upper-case letters to denote all multi-object states.
The number of elements in a set $X$, its cardinality, is denoted by $|X|$.
In the labelled RFS approach, a single-object state has the form $(x, \ell)$, where $\ell$ is a uniquely identifying track label.
We use bold letters to denote labelled single-object or multi-object states and the densities defined on them ($\mathbf{x}$, $\mathbf{X}$, and $\boldsymbol{\pi}$ respectively) to distinguish them from their unlabelled counterparts.

In the following subsections, we first define labelled RFSs and then review the two specific forms of labelled RFSs most important to this paper: The labelled and generalised labelled multi-Bernoulli RFSs.

\subsection{Labelled random finite sets}
\label{subsection:labelled_rfs_def}
An RFS is a finite-set-valued random variable.
For a given RFS, the number of points in the set is random, as are the points themselves, and they have no defined ordering within the set \cite{mahler_phd_2003}.
In this paper, we make use of finite set statistics' (FISST) notion of integration and density \cite{mahler_book_2007, mahler_book_2014}; however, for simplicity, we do not distinguish between a FISST density and a conventional probability density.

To estimate object identity, an object state $x \in \mathbb{X}$ is augmented with a label $\ell \in \mathbb{L}$, where $\mathbb{L}$ is a discrete label space, and it is represented by the tuple $\mathbf{x} = (x, \ell)$.
A labelled RFS is an RFS on $\mathbb{X} \times \mathbb{L}$, where each element in a realisation has a unique label \cite{vo_vo_lrfs_2013}.
The labels of a realisation $\mathbf{X}$ are given by 
\begin{align}
\mathcal{L} (\mathbf{X}) \triangleq \{ \ell : (x, \ell) \in \mathbf{X} \} .
\end{align} 
Since the labels of $\mathbf{X}$ must be distinct, we have $|\mathcal{L} (\mathbf{X})| = |\mathbf{X}|$ and we define the distinct label indicator
\begin{align}
\Delta (\mathbf{X}) \triangleq \delta_{|\mathbf{X}|} (|\mathcal{L} (\mathbf{X})|)
\end{align}
to ensure this condition, where 
\begin{align}
\delta_{Y} (X) \triangleq \begin{cases}
1, \ \text{if} \ X = Y \\
0, \ \text{if otherwise}
\end{cases}
\end{align}
is a generalised Kronecker delta for arbitrary inputs $X$ and $Y$.

\subsubsection{Labelled multi-Bernoulli (LMB) RFS}
\label{subsubsection:lmb_rfs_def_sect}
An LMB RFS \cite{vo_vo_lmb_2014} with a finite label space $\mathbb{L}$ is described by the parameter set
\begin{align}
\boldsymbol{\pi} =  \{ (r^{(\ell)}, p^{(\ell)} ) \}_{\ell \in \mathbb{L}},
\end{align}
where an element in the set corresponds to an object with a unique label $\ell$, an existence probability $r^{(\ell)}$, and a spatial probability distribution $p^{(\ell)} (x)$.
An LMB RFS is distributed according to
\begin{align}
\boldsymbol{\pi} ( \mathbf{X}  ) &= \Delta(\mathbf{X}) w(\mathcal{L} (\mathbf{X})) \prod_{ (x, \ell) \in \mathbf{X} } p^{(\ell)} (x), \label{eqn:lmb_density_def}
\end{align}
where the LMB weight is defined as
\begin{align}
w(L) \triangleq \prod_{i \in \mathbb{L}} (1 - r^{(i)})  \prod_{\ell \in L} \frac{1_{\mathbb{L}} (\ell) r^{(\ell)}}{1 - r^{(\ell)}} \label{eqn:lmb_weight_def},
\end{align}
such that $\sum_{L \subseteq \mathbb{L} } w(L) = 1.$
In Equation \ref{eqn:lmb_weight_def}, we make use of a generalised indicator function
\begin{align}
1_{Y} (X) \triangleq \begin{cases}
1, \ \text{if} \ X \subseteq Y \\
0, \ \text{if otherwise}
\end{cases},
\end{align}
which is defined  for arbitrary input sets $X$ and $Y$. 
Throughout this paper, we abbreviate an LMB density using its parameter set, i.e. $\boldsymbol{\pi} =  \{ (r^{(\ell)}, p^{(\ell)} ) \}_{\ell \in \mathbb{L}}.$

\subsubsection{Generalised labelled multi-Bernoulli (GLMB) RFS}
\label{subsubsection:glmb_rfs_def_sect}
In this paper, we only consider a specific form of the GLMB distribution, the LMB mixture distribution.
This GLMB form arises in the LMB filter when applying the multi-object likelihood function to an LMB prior, due to the data association uncertainty inherent in the multi-object measurement model.
A GLMB RFS \cite{vo_vo_lmb_2014} with the finite label space $\mathbb{L}$ has the density
\begin{align}
\boldsymbol{\pi}(\mathbf{X}) = \Delta(\mathbf{X}) \sum_{c \in \mathbb{C}} \sigma^{(c)}  w^{(c)} (\mathcal{L}(\mathbf{X})) \prod_{ (x, \ell) \in \mathbf{X} } p^{(c, \ell)} (x) .
\end{align}
As before, each object is uniquely associated with a label $\ell \in \mathbb{L}$.
The discrete set $\mathbb{C}$ indexes an LMB component, and an object has the existence probability $r^{(c, \ell)}$ and spatial distribution $p^{(c, \ell)} (x)$ for each $c \in \mathbb{C}$.
Each component in the mixture has the LMB weight
\begin{align}
w^{(c)} (L) = \prod_{i \in \mathbb{L}} (1 - r^{(c, i)})  \prod_{\ell \in L} \frac{1_{\mathbb{L}} (\ell) r^{(c, \ell)}}{1 - r^{(c, \ell)}}, \label{eqn:lmbm_weight_def}
\end{align}
and the additional non-negative weight $\sigma^{(c)}$, where $\sum_{c \in \mathbb{C}} \sigma^{(c)} = 1. $
This GLMB form's labelled probability hypothesis density (PHD) and expected cardinality are respectively given by
\begin{align}
v(x, \ell) &= 1_{\mathbb{L}} (\ell) \sum_{c \in \mathbb{C}} \sigma^{(c)} r^{(c, \ell)} p^{(c, \ell)} (x), \label{eqn:reduced_phd} \\
\overline{n} &= \sum_{\ell \in \mathbb{L}} \sum_{c \in \mathbb{C}} \sigma^{(c)} r^{(c, \ell)}. \label{eqn:reduced_card}
\end{align}

\subsection{The multi-sensor multi-object Bayes filter}
\label{subsection:multi_object_bayes_filter}
In this section, we review the multi-sensor multi-object Bayes filter, which is the theoretical underpinning of all RFS-based multi-object filtering and data fusion.
It is a direct generalisation of the single-object Bayes filter defined on a space of RFSs \cite{mahler_phd_2003}.
In general, the multi-sensor multi-object Bayes filter is computationally intractable and, in this paper, we develop methods of efficiently approximating it.
We assume the objects obey the standard multi-object motion model and they are partially observed by multiple independent sensors, each obeying the standard multi-object measurement model.
The multi-object state is modelled as a labelled RFS on $\mathbb{X} \times \mathbb{L}$.
The measurements collected from sensor $i$ are a partial observation of the multi-object state $\mathbf{X}$ contaminated with noise and clutter.
The measurements are modelled as the RFS $Z^{(i)} \subset \mathbb{Z}$, where $\mathbb{Z}$ is the multi-object measurement space.

If we have $S$ independent sensors, then at the current time-step the multi-object posterior is given by \cite{vo_vo_lrfs_2013, vo_multisensor_2017}
\begin{align}
\boldsymbol{\pi} ( \mathbf{X}_{+} | Z) = \dfrac{g(Z | \mathbf{X}_{+} ) \boldsymbol{\pi}_{+} (\mathbf{X}_{+}) }{\int g(Z | \mathbf{X}_{+} ) \boldsymbol{\pi}_{+} (\mathbf{X}_{+}) \delta \mathbf{X}_{+} } , \label{eqn:multi_sensor_object_bayes_def}
\end{align}
where $Z \triangleq Z^{(1)}, \dots, Z^{(S)}$, the likelihood function is a product of each sensor's likelihood function
\begin{align}
g (Z | \mathbf{X}_{+}) = \prod_{i=1}^{S} g^{(i)} (Z^{(i)} | \mathbf{X}_{+}) , \label{eqn:likelihood_bayes_prod_def}
\end{align}
and $\boldsymbol{\pi}_{+} (\mathbf{X}_{+})$ is the multi-object prior.
As it is defined using RFSs, the posterior contains all information about both the number of objects and their states.
The multi-object likelihood function $g^{(i)} (Z^{(i)} | \mathbf{X}_{+})$ models sensor $i$'s stochastic measurement process, accounting for measurement noise, data association uncertainty and clutter-generated measurements.
The multi-object prior is computed using the multi-object Chapman-Kolmogorov equation
\begin{align}
\boldsymbol{\pi}_{+} (\mathbf{X}_{+}) = \int \mathbf{f} (\mathbf{X}_{+} | \mathbf{X}) \boldsymbol{\pi} (\mathbf{X}) \delta \mathbf{X} \text{,} \label{eqn:basic_chapman_kolmo_def}
\end{align}
where the set integral is defined as
\begin{align}
\int \mathbf{f}(\mathbf{X}) \delta \mathbf{X} \triangleq \sum_{n \geq 0} \frac{1}{n!} \sum_{(\ell_{1}, \dots, \ell_{n}) \in \mathbb{L}^{n}} \int_{\mathbb{X}^{n}} \mathbf{f}( \{ (x_{1}, \ell_{1}), \dots, (x_{n}, \ell_{n}) \} ) d (x_{1}, \dots, x_{n}),
\end{align}
$\mathbf{f} (\mathbf{X}_{+} | \mathbf{X})$ is the multi-object Markov density and $\boldsymbol{\pi} (\mathbf{X})$ is the multi-object posterior at the previous time-step. The multi-object Markov density provides an equivalent description of the motion model, describing object birth, death and motion.
Together, Equations \ref{eqn:multi_sensor_object_bayes_def} and \ref{eqn:basic_chapman_kolmo_def} form a recursive definition which is repeated at every time-step.

In the following subsections, we discuss the standard multi-object Markov density $\mathbf{f} (\mathbf{X}_{+} | \mathbf{X})$ and single-sensor likelihood function $g^{(i)}(Z^{(i)} | \mathbf{X}_{+})$ used in Equations \ref{eqn:basic_chapman_kolmo_def} and \ref{eqn:likelihood_bayes_prod_def} respectively. 

\subsubsection{Multi-object Markov density}
\label{subsubsection:multi_object_markov_density}
We now discuss the two major components of the multi-object Markov density $\mathbf{f} (\mathbf{X}_{+} | \mathbf{X})$: the density describing the death and motion of objects in the previous multi-object state, and the density describing the appearance of new objects.
Given a valid state set $\mathbf{X}$ with distinct labels at the previous time-step, then each object $( x, \ell) \in \mathbf{X}$ either survives to the current time-step with probability $P_{S} (x, \ell)$ and state $(x_{+}, \ell_{+})$ distributed according to the Markov density $f_{+} (x_{+} | x, \ell)$, or it dies with probability $1 - P_{S} (x, \ell)$.
The objects' state transitions are assumed to be independent of one another, given a valid $\mathbf{X}$.
Therefore, at the current time-step, the set of surviving objects $\mathbf{W}$ is distributed according to the LMB distribution \cite{vo_vo_lrfs_2013, vo_vo_lmb_2014}
\begin{align}
\mathbf{f}_{S} (\mathbf{W} | \mathbf{X}) = \Delta(\mathbf{W}) 1_{\mathcal{L} (\mathbf{X})} (\mathcal{L} (\mathbf{W})) \prod_{ (x, \ell) \in \mathbf{X} } \Phi (\mathbf{W}; x, \ell) , \label{eqn:multi_Bernoulli_survival}
\end{align}
where the single-object state transition model is given by
\begin{align}
\Phi(\mathbf{W}; x, \ell) = \begin{cases}
P_{S} (x, \ell) f_{+} ( x_{+} | x, \ell), \ & \text{if} \ \ell \in \mathcal{L} (\mathbf{W}) \\
1 - P_{S} (x, \ell), \ & \text{otherwise}
\end{cases}.
\end{align}

Object appearance is assumed to occur independently of the existing objects.
In this paper, we only consider a multi-Bernoulli appearance model.
The set of objects $\mathbf{Y}$ born at the current time-step is distributed according to an LMB density with the parameter set \cite{vo_vo_lrfs_2013, vo_vo_lmb_2014}
\begin{align}
\boldsymbol{\pi}_{B} = \{ ( r_{B}^{(\ell)}, p_{B}^{(\ell)} )  \}_{\ell \in \mathbb{B}}, \label{eqn:lmb_birth_density_def}
\end{align}
where $\mathbb{B}$ is a finite label space.
The label space at the current time-step is given by the disjoint union $\mathbb{L}_{+} = \mathbb{B} \uplus  \mathbb{L}$, where $\mathbb{L}$ is the previous label space.
To ensure $\mathbb{L}$ and $\mathbb{B}$ are disjoint, we follow Vo and Vo's labelling convention \cite{vo_vo_lrfs_2013}:
Each object is identified by the unique label $\ell = (k, i)$, where $k$ is the time of birth and $i \in \mathbb{N}$ is a unique index used to ensure objects with the same birth time are distinguishable.

Since the objects move, appear and disappear independently of one another, the multi-object state at the current time-step is given by $\mathbf{X}_{+} = \mathbf{W} \uplus \mathbf{Y}$ and it is distributed according to the LMB distribution \cite{vo_vo_lrfs_2013, vo_vo_lmb_2014}:
\begin{align}
\mathbf{f} (\mathbf{W} \uplus \mathbf{Y} | \mathbf{X}) = \boldsymbol{\pi}_{B} ( \mathbf{Y} ) \mathbf{f}_{S} ( \mathbf{W} | \mathbf{X})  \text{.}
\end{align}

\subsubsection{Multi-object likelihood function}
\label{subsubsection:multi_object_likelihood}
Since measurements can be generated by either objects or by clutter, the standard multi-object likelihood function $g^{(i)} (Z | \mathbf{X}_{+})$ for sensor $i$ is constructed by combining the measurement models for both.
If we have a valid state set $\mathbf{X}_{+}$, then each object $\mathbf{x} \in \mathbf{X}_{+}$  has a probability $P_{D} (\mathbf{x})$ of its measurement $z$ being detected, and a probability $1 - P_{D} (\mathbf{x})$ of $z$ going undetected. Conditioned on $\mathbf{X}_{+}$, we assume the object-generated measurements are independent of both one another and the clutter-generated measurements.
The set of object-generated measurements therefore has a multi-Bernoulli distribution with probability density \cite{vo_vo_lrfs_2013}
\begin{align}
\pi_{D} (W | \mathbf{X}_{+}) = \{ ( P_{D} (\mathbf{x}), g( \cdot | \mathbf{x}) ) : \mathbf{x} \in \mathbf{X}_{+} \} (W).
\end{align}
We assume the clutter measurements are modelled by a Poisson RFS distributed according to
\begin{align}
\pi_{K} (Y) = e^{\langle -\kappa, 1 \rangle} \prod_{y \in Y} \kappa(y),
\end{align}
where $\kappa (\cdot)$ is the clutter intensity function and we abbreviate the inner product of functions $f$ and $g$ using
\begin{align}
\langle f, g \rangle \triangleq \int f(x) g (x) d x. \label{eqn:inner_product_def}
\end{align}
Typically, $\kappa(y) = \lambda c(y)$, where $\lambda$ is the expected number of clutter returns per time-step and $c(y)$ is uniform distribution over surveillance region \cite[pp. 410]{mahler_book_2007}.

The multi-object measurement $Z^{(i)}$ is the disjoint union of the clutter- and object-generated measurements, $Z^{(i)} = W \uplus Y$, and the multi-object likelihood function is given by \cite{vo_multisensor_2017}
\begin{align}
g^{(i)} (Z^{(i)} | \mathbf{X}_{+}) &= \pi_{K} (Z^{(i)}) \sum_{\omega \in \Omega} \prod_{(\mathbf{x}, \ell) \in \mathbf{X}_{+}} L(z_{\omega(\ell)}|x, \ell). \label{eqn:multi_object_likelihood_function}
\end{align}
Here $\Omega$ is a space of association functions $\omega : \mathcal{L} (\mathbf{X}_{+}) \rightarrow \{0, 1, \dots, |Z^{(i)}| \}$, such that $\omega(i) = \omega(j) > 0$ implies $i = j$, and we use the likelihood ratios
\begin{align}
L (z_{j}|x, \ell) &\triangleq \begin{cases}
1 - P_{D} (x, \ell), & \text{if} \ j = 0 \\
\dfrac{P_{D} (x, \ell) g(z_{j} | x, \ell) }{ \kappa (z_{j})  }, & \text{if} \ j > 0
\end{cases}. \label{eqn:multi_object_likelihood_ratios}
\end{align}
We use association functions to index association events, the unique joint assignment of measurements to objects.
In this case, if $\omega(\ell) = 0$, then object $\ell$ missed its detection. If $\omega(\ell) = j$, then object $\ell$ generated measurement $j$ with likelihood $g(z_{j} | x, \ell)$.

\subsection{The LMB filter}
The LMB filter is an efficient approximation of the labelled multi-object Bayes filter \cite{vo_vo_lmb_2014}, which approximates the multi-object Bayes filter's exact posterior distribution using a single LMB distribution.
In the following sections, we briefly review the prediction and measurement update steps of the single-sensor LMB filter.
This section's results form the basis of both the PU- and GA-LMB filters in Sections \ref{section:pu_lmb} and \ref{section:ci_lmb} respectively.

\subsubsection{Prediction}
\label{subsection:lmb_filter_pred}
In the LMB filter's prediction step, we determine the prior distribution on the current multi-object state, accounting for object appearance, transition and death.
Assuming a multi-Bernoulli object appearance model, the LMB density is closed under the Chapman-Kolmogorov equation \cite{vo_vo_lmb_2014}.
By our assumptions, the previous time-step's posterior LMB distribution is parameterised by
\begin{align}
\boldsymbol{\pi} = \{ ( r^{(\ell)}, \ p^{(\ell)} )  \}_{\ell \in \mathbb{L}},
\end{align}
and the birth LMB density is given in Equation~\ref{eqn:lmb_birth_density_def}.
The multi-object Chapman-Kolmogorov equation (Equation \ref{eqn:basic_chapman_kolmo_def}) yields an LMB prior on the current multi-object state which is parameterised by \cite{vo_vo_lmb_2014}
\begin{align}
\boldsymbol{\pi}_{+} =  \{ ( r_{+, S}^{(\ell)}, \ p_{+}^{(\ell)} ) \}_{\ell \in \mathbb{L}} \cup \{  (r_{B}^{(\ell)}, \ p_{B}^{(\ell)} ) \}_{\ell \in \mathbb{B}} .\label{eqn:prior_lmb_density_pred}
\end{align}
This is simply the union of the surviving objects' and newly appearing objects' LMB parameter sets, where a surviving object's prior existence probability and spatial distribution are respectively given by
\begin{align}
r_{+, S}^{(\ell)} &\triangleq \eta_{S}(\ell) r^{(\ell)} , \\
p_{+, S}^{(\ell)} (x) &\triangleq \dfrac{ \langle P_{S} (\cdot, \ell), f_{+}(x | \cdot, \ell) p^{(\ell)} (\cdot) \rangle  }{ \eta_{S} (\ell) } , \label{eqn:single_object_transition} \\
\eta_{S} (\ell) &\triangleq \langle P_{S} (\cdot, \ell), p^{(\ell)}(\cdot) \rangle.
\end{align}

\subsubsection{Measurement update}
\label{subsubsection:measurement_update}
The LMB distribution is not a conjugate prior for the standard multi-object likelihood function \cite{vo_vo_lmb_2014}, and, to avoid propagating a more complex distribution, the LMB filter approximates GLMB posterior distribution using a single LMB component.
If we have an LMB prior parameterised by the set $\boldsymbol{\pi}_{+} = \{ ( r_{+}^{(\ell)}, p_{+}^{(\ell)} ) \}_{\ell \in \mathbb{L}_{+}} , $ then the posterior density is the GLMB \cite{mahler_labelled_cphd_2017}
\begin{align}
\boldsymbol{\pi} (\mathbf{X}_{+} | Z^{(i)}) = \Delta (\mathbf{X}_{+}) \sum_{\theta \in \Theta} \sigma^{(\theta)} w^{(\theta)} (\mathcal{L} (\mathbf{X}_{+})) \prod_{(x, \ell) \in \mathbf{X}_{+}}p^{(\theta, \ell)} (x) . \label{eqn:glmb_posterior}
\end{align}
Similar to Section \ref{subsubsection:multi_object_likelihood}, $\Theta$ is a space of association functions $\theta : \mathbb{L}_{+} \rightarrow \{0, 1, \dots, |Z^{(i)}| \}$, such that $\theta(i) = \theta(j) > 0$ implies $i = j$.
The function $\theta$ indexes an association event, or hypothesis. Under hypothesis $\theta$, object $\ell$ has the respective posterior existence probability and spatial distribution 
\begin{align}
r^{(\theta, \ell)} &\triangleq \begin{cases}
\dfrac{r_{+}^{(\ell)} [ 1 -  \langle P_{D} (\cdot, \ell), p_{+}^{(\ell)} (\cdot) \rangle ]}{1 - r_{+}^{(\ell)} \langle P_{D} (\cdot, \ell), p_{+}^{(\ell)} (\cdot) \rangle}, & \text{if} \ \theta(\ell) = 0 \\
1, & \text{if} \ \theta(\ell) > 0 
\end{cases} ,  \label{eqn:post_exist_lmb} \\
p^{(\theta, \ell)} (x) &\triangleq \begin{cases}
\dfrac{[1 - P_{D} (x, \ell)] p_{+} (x, \ell) }{1 - \langle P_{D} (\cdot, \ell), p_{+}^{(\ell)} (\cdot) \rangle}, & \text{if} \ \theta(\ell) = 0 \\
\dfrac{P_{D} (x, \ell) g (z_{\theta(\ell)} | x , \ell) p_{+}^{(\ell)} (x) }{ \langle P_{D} (\cdot, \ell), g (z_{\theta(\ell)} | \cdot, \ell ) p_{+}^{(\ell)} (\cdot) \rangle  }, & \text{if} \ \theta(\ell) > 0 
\end{cases} . \label{eqn:post_spatial}
\end{align}
If the object misses its detection, such that $\theta(\ell) = 0$, then it is possible that the object does not exist and its spatial distribution equals its prior.
If the object is assigned measurement $k$, such that $\theta(\ell) = k$, then it must exist and its prior spatial distribution is updated using measurement $k$.
Hypothesis $\theta$ has the LMB weight
\begin{align}
w^{(\theta)} (L) &\triangleq \prod_{i \in \mathbb{L}_{+}} (1 - r^{(\theta, i)}) \prod_{\ell \in L} \frac{1_{\mathbb{L}_{+}} (\ell) r^{(\theta, \ell)}}{1 - r^{(\theta, \ell)}},
\end{align}
and the additional weight
\begin{align}
\sigma^{(\theta)} &\triangleq \frac{R^{(\theta)}}{\sum_{\theta \in \Theta}  R^{(\theta)} }, \\
R^{(\theta)} &\triangleq \prod_{\ell: \theta(\ell) > 0} \dfrac{r_{+}^{(\ell)} \langle P_{D} (\cdot, \ell), g (z_{\theta(\ell)} | \cdot, \ell) p_{+}^{(\ell)} (\cdot) \rangle}{ \kappa(z_{\theta(\ell)}) [ 1 - r_{+}^{(\ell)} \langle P_{D} (\cdot, \ell), p_{+}^{(\ell)} (\cdot) \rangle ]}.
\end{align}
The GLMB posterior (Equation \ref{eqn:glmb_posterior}) is now fully specified.

To approximate the posterior GLMB using an LMB, we use an LMB distribution that matches the GLMB's first moment, as it minimises the multi-object Kullback-Leibler divergence between the two distributions \cite{vo_vo_lmb_2014}.
To do this, we compute an object's labelled PHD using Equation \ref{eqn:reduced_phd},
\begin{align}
v(x, \ell) &= 1_{\mathbb{L}_{+}} (\ell) \sum_{\theta \in \Theta} \sigma^{(\theta)} r^{(\theta, \ell)} p^{(\theta, \ell)}(x),
\end{align}
and use the PHD to approximate an object's existence probability and spatial distribution:
\begin{align}
r^{(\ell)} &= \int v(x, \ell) d x = 1_{\mathbb{L}_{+}} (\ell) \sum_{\theta \in \Theta}  \sigma^{(\theta)} r^{(\theta, \ell)}, \\
p^{(\ell)}(x) &= 1_{\mathbb{L}_{+}} (\ell) \frac{1}{r^{(\ell)}} \sum_{\theta \in \Theta} \sigma^{(\theta)} r^{(\theta, \ell)} p^{(\theta, \ell)} (x).
\end{align}
Put simply, each of an object's approximate parameters is the weighted average of its corresponding posterior parameter under each hypothesis.
The approximate posterior LMB is represented by the parameter set
\begin{align}
\boldsymbol{\pi} = \{ ( r^{(\ell)}, p^{(\ell)} ) \}_{\ell \in \mathbb{L}_{+}} , \label{eqn:posterior_single_sensor_lmb}
\end{align}
and with that the LMB filter is fully described.

\section{Efficient approximation of the LMB filter}
\label{section:efficient_approximation_lbp_lmb}
In this section, we present a method of efficiently approximating the LMB filter's measurement update using loopy belief propagation (LBP).
The LMB filter can be implemented using a Gibbs sampler, with a quadratic complexity in object number and linear complexity in measurement and sample number \cite{Reuter_gibbs_2017}.
In contrast, the LBP approach we present here has linear complexity in object number.
It is a modification of Williams et al.'s LBP approach to data association \cite{williams_lau_lbp_2010}, defined on a cluster graph, rather than a factor graph.
Variations of Williams et al.'s algorithm, which yield the same results as our approach, have already been successfully used to approximate the LMB filter \cite{yang_wang_lbp_2018, meyer_2019_lmb_lbp}.

\subsection{Probabilistic graphical models and belief propagation}
In this subsection, we introduce our chosen probabilistic graphical model (PGM), the cluster graph, and define a belief propagation message passing algorithm on it.
PGMs provide powerful methods of representing and manipulating complex statistical models.
Graphs are used to model variable interaction, and the variables' accompanying joint distribution is given a modular factorisation that coincides with their interaction.
This sparse representation makes inference not only computationally tractable, but efficient.

A PGM \cite{koller_pgms_2009} is a factored representation of a joint distribution defined over the variable set $X$.
The distribution is parametrised by the factor set 
$\Phi = \{ \phi_{1}(X_{1}), \dots, \phi_{n} (X_{n}) \},$ where each factor $\phi_{i}(X_{i})$ is defined over the variable subset $X_{i} \subseteq X$.
The joint distribution is given by 
\begin{align}
p(X) = \frac{1}{\eta} \prod_{i=1}^{n} \phi_{i} (X_{i}),
\end{align}
where $\eta$ is a normalising constant.
A cluster graph \cite[pp. 346]{koller_pgms_2009} for the factor set $\Phi$ over $X$ is an undirected graph, where each node, or cluster, is associated with a unique subset of variables $C_{j} \subseteq X$, and each factor $\phi_{i}(X_{i}) \in \Phi$ is associated with a single cluster $C_{j}$ such that $X_{i} \subseteq C_{j}$.
The edge between an adjacent pair of clusters is associated with a separator set (sepset) $S_{j, k} \subseteq C_{j} \cap C_{k}$.

The purpose of message passing is to approximate the marginal distributions of the cluster variables.
On a cluster graph, a message sent between two neighbouring clusters $C_{i}$ and $C_{j}$ is a factor defined on their sepset $S_{j, k}$.
In belief propagation (BP) message passing, a message is defined as \cite[pp. 352]{koller_pgms_2009}:
\begin{align}
\mu_{i \rightarrow j} (S_{i, j}) \triangleq \sum_{C_{i} - S_{i, j}} \psi_{i} (C_{i}) \prod_{k \in \Lambda_{i} - \{ j \} } \mu_{k \rightarrow i} (S_{k, i}), \label{eqn:sum_product_def}
\end{align}
where $\Lambda_{i}$ is the index set of all $C_{i}$'s adjacent clusters and $\psi_{i} (C_{i})$ is the cluster potential -- the product of all factors $\phi_{k} (X_{k}) \in \Phi$ assigned to $C_{i}$.
When dealing with continuous variables, summation is replaced with integration. Once cluster $C_{j}$ has received all incoming messages, it can express all information it holds over its domain as the belief distribution
\begin{align}
\beta (C_{j}) = \psi_{j} (C_{j}) \prod_{k \in \Lambda_{j}} \mu_{k \rightarrow j} (S_{k, j}). \label{eqn:cluster_belief_def}
\end{align}
Messages are passed iteratively between clusters until convergence, when all clusters agree on their shared marginal beliefs, such that \cite[pp. 358]{koller_pgms_2009}
\begin{align}
\sum_{C_{i} - S_{i, j}} \beta(C_{i}) = \sum_{C_{j} - S_{i, j}} \beta(C_{j})
\end{align}
for all pairs $C_{i}$ and $C_{j}$ of adjacent clusters.
When applied to a loopy graph, BP is not guaranteed to converge to the correct answer, or even converge at all.
However, the LBP algorithm presented in this section reduces to a system of fixed-point equations, and equations of their form are guaranteed to converge \cite{williams_lau_con_2010}.
The resulting approximate marginal association probabilities are of sufficient accuracy to implement a robust LMB filter \cite{yang_wang_lbp_2018, meyer_2019_lmb_lbp}.

\subsection{Data association PGM}
\label{subsection:approximate_implemtation}
In this subsection, we first provide a cluster graph formulation of the LMB filter's data association problem, which we believe allows for a concise, transparent encoding of the problem.
We then use LBP to approximate the filter's measurement update.

\subsubsection{Cluster graph construction}
If the number of predicted objects is given by $n = |\mathbb{L}_{+}|$ and a single sensor collects $m$ measurements, then the GLMB posterior distribution in Equation \ref{eqn:glmb_posterior} enumerates $ \sum_{k \geq 0} k! \binom{n}{k} \binom{m}{k} $ joint association events.
If $n=20$ and $m=20$, then there are approximately $1.38 \times 10^{12}$ joint association events.
Determining the exact LMB posterior distribution using this GLMB distribution is prohibitively expensive, even for a modest $n$ and $m$, and we must resort to approximation.

Given the prior parameter set $\{ (r_{+}^{(\ell_{i})}, p_{+}^{(\ell_{i})} ) \}_{i=1}^{n}$, we must find an approximate representation of the posterior LMB density for which we can then define a factor set and cluster graph.
This approximate distribution takes a simpler form that allows for inexpensive inference.
Our approximation does not use a FISST density, but rather a conventional probability distribution defined on a vectorisation of the multi-object state.
First, we model object $i$'s existence using the binary variable $e_{i}$, where $e_{i} = 1$ indicates the object exists, and $e_{i} = 0$ indicates that it does not.
To describe the association events, we define $m+1$ binary association variables, $a_{i}^{0:m} \triangleq a_{i}^{0}, a_{i}^{1}, \dots, a_{i}^{m},$ for each object $i$.
Here $a_{i}^{0} = 1$ indicates object $i$ missed its detection, while $a_{i}^{0} = 0$ indicates that it did not.
For $j>0$,  $a_{i}^{j} = 1$ indicates object $i$ generated $z_{j}$, while $a_{i}^{j} = 0$ indicates it did not.
In Table \ref{table:initial_target_belief}, 
\iftrue
\begin{table}
	\centering
	\begin{tabular}{|cccccc|l|} \hline
    $e_{i}$ & $a_{i}^{0}$ & $a_{i}^{1}$ & $a_{i}^{2}$ & $\cdots$ & $a_{i}^{m}$ & $\phi_{T_{i}} (x_{i}, z_{0:m}, a_{i}^{0:m}, e_{i})$ \\ \hline
    0 & 0 & 0 & 0 & $\cdots$ & 0 & $ (1 - r_{+}^{(\ell_{i})})  p_{+}^{(\ell_{i})}(x_{i}) $ \\
    1 & 1 & 0 & 0 & $\cdots$ & 0 & $  r_{+}^{(\ell_{i})} L (z_{0} | x_{i}, \ell_{i}) p_{+}^{(\ell_{i})}(x_{i}) $ \\
    1 & 0 & 1 & 0 & $\cdots$ & 0 & $  r_{+}^{(\ell_{i})} L (z_{1} | x_{i}, \ell_{i})  p_{+}^{(\ell_{i})}(x_{i}) $ \\
    1 & 0 & 0 & 1 & $\cdots$ & 0 & $  r_{+}^{(\ell_{i})} L (z_{2} | x_{i}, \ell_{i})  p_{+}^{(\ell_{i})}(x_{i}) $ \\
    $\vdots$ & $\vdots$ & $\vdots$ & $\vdots$ & $\ddots$ & $\vdots$ & $\vdots$ \\ 
    $1$ & 0 & 0 & 0 & $\cdots$ & 1 & $  r_{+}^{(\ell_{i})} L (z_{m} | x_{i}, \ell_{i}) p_{+}^{(\ell_{i})}(x_{i}) $ \\  \hline
	\multicolumn{6}{|c|}{Elsewhere} & 0 \\ \hline
	\end{tabular}
	\caption{A factor enforcing the constraint that an object $i$ can generate at most one of the $m$ measurements.
	The likelihood ratios are defined in Equation \ref{eqn:multi_object_likelihood_ratios}, and $z_{0:m} \triangleq z_{0}, z_{1}, \dots, z_{m}$. Here $z_{0}$ is a notational convenience that indicates an object missed its detection.}
	\label{table:initial_target_belief}
\end{table}
\fi
we define a factor which models that object $i$ generates at most one measurement and lists the likelihoods of these events.
If $e_{i} = 1$, then at most one association variable may equal one in each entry in Table \ref{table:initial_target_belief}, while the rest must equal zero.
When $e_{i} = 0$, all the association variables must equal zero.
We abbreviate each table entry using the likelihood ratios defined in Equation \ref{eqn:multi_object_likelihood_ratios}, and we also use the notation $z_{0:m} \triangleq z_{0}, z_{1}, \dots, z_{m}$, where $z_{0}$ is a notational convenience that indicates an object missed its detection.
To ensure Table \ref{table:initial_target_belief} is a valid joint measure, an object is assigned its prior spatial distribution when it does not exist, but this distribution plays no role in the resulting inference.
Finally, we define a factor in Table \ref{table:initial_measurement_belief} 
\iftrue
\begin{table}
	\centering
     \begin{tabular}{|cccc|c|} \hline
	 $a_{1}^{j}$ & $a_{2}^{j}$ & $\cdots$ & $a_{n}^{j}$ & $\phi_{M_{j}} (a_{1:n}^{j})$ \\ \hline
	 0 & 0 & $\cdots$  & 0 & 1 \\
	 1 & 0 & $\cdots$  & 0 & 1 \\
	 0 & 1 & $\cdots$  & 0 & 1 \\
	 $\vdots$ & $\vdots$ & $\ddots$ & $\vdots$ & 1 \\
	 0 & 0 & $\cdots$  & 1 & 1 \\ \hline
	 \multicolumn{4}{|c|}{Elsewhere} & 0 \\ \hline
	\end{tabular}
    \caption {A factor enforcing the constraint that measurement $j$ is generated by at most one of the $n$ objects.}
    \label{table:initial_measurement_belief}
\end{table}
\fi
which models that every measurement is generated by at most one object, where $a_{1:n}^{j} \triangleq a_{1}^{j}, \dots, a_{n}^{j}$.
Again, for each entry in Table~\ref{table:initial_measurement_belief}, at most one association variable may equal one, while the rest must equal zero.
Using Tables~\ref{table:initial_target_belief} and~\ref{table:initial_measurement_belief}, we approximate the posterior LMB distribution as
\begin{align}
\pi (x_{1:n}, a_{1:n}, e_{1:n} \big| z_{0:m}) = \frac{1}{\eta} \prod_{i = 1}^{n} \phi_{T_{i}} (x_{i}, z_{0:m}, a_{i}^{0:m}, e_{i}) \prod_{j=1}^{m} \phi_{M_{j}} (a_{1:n}^{j}), \label{eqn:approx_lmb_posterior}
\end{align}
where $x_{1:n} \triangleq x_{1}, \dots, x_{n}$, $a_{1:n} \triangleq a_{1}^{0:m}, \dots, a_{n}^{0:m}$, $e_{1:n} \triangleq e_{1}, \dots e_{n}$, and $\eta$ is a normalising constant.
In principle, we could approximate each object's posterior existence probability and spatial distribution using
\begin{align}
p^{(\ell_{i})} (x_{i}, e_{i}) &= \sum_{\{ a_{1:n} \}, \{ e_{1:n} \} - \{ e_{i} \} } \int \pi (x_{1:n}, a_{1:n}, e_{1:n} \big| z_{0:m})  d (x_{1} \dots x_{i-1} x_{i+1} \dots  x_{n}), \\
r^{(\ell_{i})} &\approx \int p^{(\ell_{i})} (x_{i}, e_{i} = 1) dx_{i}, \label{eqn:approx_post_existence}  \\
p^{(\ell_{i})}(x_{i}) &\approx \frac{1}{r^{(\ell_{i})}}   p^{(\ell_{i})} (x_{i}, e_{i} = 1). \label{eqn:approx_post_spatial}
\end{align}
However, this is also prohibitively expensive, and we must resort to further approximation by LBP.

We now construct a cluster graph using our factor set. We define a cluster $T_{i} = \{ x_{i}, z_{0:m}, a_{i}^{0:m}, e_{i}  \}$
for each object, and a cluster  $M_{j} = \{ a_{1:n}^{j} \}$ for each measurement.
An object cluster $T_{i}$ is connected to measurement cluster $M_{j}$ through the singleton sepset $S_{i, j} = T_{i} \cap M_{j} = \{ a_{i}^{j} \} $ resulting in a bipartite graph, an example of which is given in Figure \ref{fig:targetToMeasurement}.
\iftrue 
\begin{figure}
	\centering
	\includegraphics[scale=0.75]{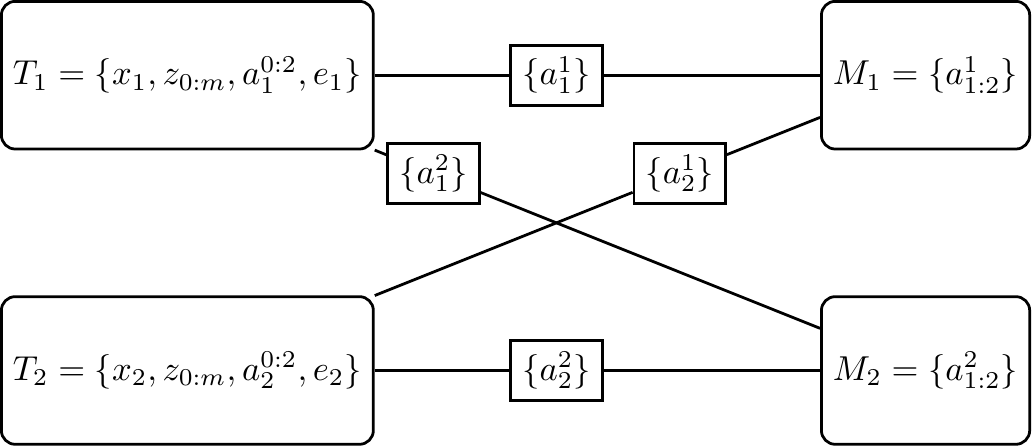}
	\caption{A bipartite cluster graph defined for the data association problem, given two objects and two measurements.
	The clusters, named $T_{i}$ and $M_{j}$, are represented by the rectangles with rounded edges, while the sepsets are represented using basic rectangles.}
	\label{fig:targetToMeasurement}
\end{figure}
\fi
For their cluster potentials, the object and measurement clusters are assigned the factors in Tables \ref{table:initial_target_belief} and \ref{table:initial_measurement_belief} respectively,
\begin{align}
\psi_{T_{i}} (x_{i}, z_{0:m}, a_{i}^{0:m}, e_{i}) &= \phi_{T_{i}} (x_{i}, z_{0:m}, a_{i}^{0:m}, e_{i}), \\
\psi_{M_{j}} (a_{1:n}^{j}) &= \phi_{M_{j}} (a_{1:n}^{j}).
\end{align}

\subsubsection{Belief propagation}
In this section, we use LBP to approximate the objects' posterior existence probabilities and spatial distributions.
Focusing on a single iteration and using Equation~\ref{eqn:sum_product_def}, the message a measurement cluster $M_{j}$ sends to an object cluster $T_{i}$ is given by
\begin{align}
\mu_{M_{j} \rightarrow T_{i}} (a_{i}^{j}) = \sum_{\{ a_{1:n}^{j} \} - \{ a_{i}^{j} \} } \psi_{M_{j}} (a_{1:n}^{j}) \prod_{\substack{k = 1, \\ k \neq i}}^{n} \mu_{ T_{k} \rightarrow M_{j} } (a_{k}^{j}), \label{eqn:meaus_to_object_first}
\end{align}
where a message received from an adjacent object cluster is given by
\begin{align} 
\mu_{ T_{k} \rightarrow M_{j} } (a_{k}^{j}) = \begin{cases}
1, \ & \ \text{if} \ a_{k}^{j} = 0   \\
\mu_{TM}^{k, j},  & \ \text{if} \ a_{k}^{j} = 1  \\
\end{cases} .
\end{align}
We normalise all messages in this subsection relative to their first entry, as it reduces the message passing to a fixed-point system.
Using Table \ref{table:initial_measurement_belief}, Equation \ref{eqn:meaus_to_object_first} reduces to
\begin{align} 
\mu_{M_{j} \rightarrow T_{i}} (a_{i}^{j}) = \begin{cases}
1, \ & \ \text{if} \ a_{i}^{j} = 0    \\
\mu_{MT}^{i, j},  & \ \text{if} \ a_{i}^{j} = 1 \\
\end{cases}, \label{eqn:mt_messages}
\end{align}
where
\begin{align}
\mu_{MT}^{i, j} = \dfrac{1}{1 + \sum\limits_{\substack{k = 1, \\ k \neq i}}^{n} \mu_{TM}^{k, j} } . \label{eqn:reduced_mt_messages}
\end{align}
If this is the first iteration of the message passing algorithm, then $\mu_{TM}^{k,j} = 1$ for all $k$ and $j$.
The message sent from an object cluster to a measurement cluster is given by
\begin{align}
\mu_{T_{i} \rightarrow M_{j}} (a_{i}^{j}) = \sum_{e_{i}, \{ a_{i}^{0:m} \} - \{ a_{i}^{j} \} } \int \psi_{T_{i}} (x_{i}, z_{0:m}, a_{i}^{0:m}, e_{i}) d x_{i} \prod_{\substack{k = 1, \\ k \neq j}}^{m} \mu_{ M_{k} \rightarrow T_{i} } (a_{i}^{k}) . \label{eqn:tm_messages}
\end{align}
After multiplying all incoming messages into Table \ref{table:initial_target_belief}, it follows that the message in Equation \ref{eqn:tm_messages} reduces to
\begin{align}
\mu_{T_{i} \rightarrow M_{j}} (a_{i}^{j}) = \begin{cases}
1, \ & \ \text{if} \ a_{i}^{j} = 0    \\
\mu_{T M}^{i, j},  & \ \text{if} \ a_{i}^{j} = 1  \\
\end{cases}, \label{eqn:final_mt_message}
\end{align}
where
\begin{align}
\mu_{TM}^{i, j} &= \dfrac{r_{+}^{(\ell_{i})}  L_{i} (z_j) }{ 1 - r_{+}^{(\ell_{i})} + r_{+}^{(\ell_{i})} \big[ L_{i} (z_0)  + \sum\limits_{\substack{k = 1, \\ k \neq j}}^{m} L_{i}(z_k) \mu_{MT}^{i, k} \big] } , \label{eqn:reduced_final_mt_message} \\
L_{i} (z_j) &= \langle L (z_{j} | \cdot \ell_{i}), p_{+}^{(\ell_{i})} (\cdot) \rangle.
\end{align}

Equations \ref{eqn:reduced_mt_messages} and \ref{eqn:reduced_final_mt_message} are equivalent in form to those found in Williams et al.'s message passing algorithm \cite{williams_lau_con_2010}.
The equations form a fixed-point iteration, which is guaranteed to converge.
Given $n$ objects, $m$ measurements and $k$ LBP iterations, the message passing can be completed in $O(k n m)$ time, and it yields each object's approximate marginal association probabilities.
Williams et al. found this LBP approach to be highly favourable in the time versus accuracy trade-off when compared to established data association methods \cite{williams_lau_2014}.
The algorithm has also been applied to the LMB filter \cite{yang_wang_lbp_2018, meyer_2019_lmb_lbp}, and it has shown to be accurate through simulation.
In contrast, a Gibbs sampler implementation takes $O(p n^2 m)$ time to approximate an object's posterior existence probability and spatial distribution \cite{meyer_2019_lmb_lbp}, where $p$ is the number of samples.
Yang et al. \cite{yang_wang_lbp_2018} compared this LBP algorithm to both a Gibbs sampler and Murty's algorithm, and the LBP approach was found to be accurate relative to its low computational cost.
Murty's algorithm resolves the data association problem by producing the set of $t$ best association events in $O(t(m + 2n)^{3})$ time.
The algorithm is typical in MOT \cite[pp. 346]{blackman1999design}, and it has been shown to be accurate in challenging scenarios \cite{cox1996efficient, cox1995ranked}.
Murty's algorithm has been used to implement both the GLMB and LMB filter \cite{vo_vo_lrfs_2013, vo_vo_lmb_2014}, and their respective performances are comparable to a Gibbs sampler implementation \cite{vo2016efficient, Reuter_gibbs_2017}.

Once the message passing has converged, an object's posterior existence probability and spatial distribution can be obtained from its belief distribution (Equation~\ref{eqn:cluster_belief_def}), and they are respectively given by
\begin{align}
r^{(\ell_{i})} &= \frac{ r_{+}^{(\ell_{i})} \big[  L_{i} (z_0) + \sum_{k=1}^{m} L_{i} (z_k) \mu_{MT}^{i, k}  \big] }{ 1 - r_{+}^{(\ell_{i})} + r_{+}^{(\ell_{i})} \big[ L_{i} (z_0)  + \sum\limits_{k = 1}^{m} L_{i}(z_k) \mu_{MT}^{i, k} \big] }, \\
p^{(\ell_{i})}(x_{i}) &= \frac{ \big[ L(z_{0} | x_{i}, \ell_{i})  + \sum\limits_{k = 1}^{m} L(z_{k} | x_{i}, \ell_{i}) \mu_{MT}^{i, k}  \big ] p_{+}^{(\ell_{i})} (x_{i}) }{ L_{i} (z_0)  + \sum\limits_{k = 1}^{m} L_{i}(z_k) \mu_{MT}^{i, k} } .
\end{align}
The results above immediately follow from Table \ref{table:initial_target_belief}, and with that, our approximation of the LMB filter's measurement update is complete.

\section{The parallel update multi-sensor LMB (PU-LMB) filter}
\label{section:pu_lmb}
In this section, we develop an efficient, approximate multi-sensor LMB filter that allows each measurement update to be computed in parallel, and the updates are sensor order invariant.
When combined with the previous section's LBP approximation, this results in a multi-sensor LMB filter with an appealingly low computational complexity.
Our approach is based on the product multi-sensor PHD filter's approximation \cite{mahler_parallel_phd_2010}, and it requires strong simplifying assumptions: 
The prior for the measurement update must be an LMB distribution and every updated distribution produced by a sensor's measurement update is approximated as an LMB distribution.

First, we derive the parallel update LMB (PU-LMB) filter equations, we then discuss some practical concerns, focusing on a Gaussian mixture (GM) implementation.
\subsection{Derivation}
We assume there are $S$ independent sensors, each obeying the standard multi-object measurement model.
The multi-sensor, multi-object Bayes filter posterior distribution (Equation \ref{eqn:multi_sensor_object_bayes_def}) can be rearranged as follows:
\begin{align}
\boldsymbol{\pi} (\mathbf{X} | Z) = \dfrac{ \big( \boldsymbol{\pi}_{+} (\mathbf{X}) \big)^{1 - S} \prod_{i=1}^{S} \boldsymbol{\pi}^{(i)} (\mathbf{X} | Z^{(i)})}{\int \big( \boldsymbol{\pi}_{+} (\mathbf{X}) \big)^{1 - S} \prod_{i=1}^{S} \boldsymbol{\pi}^{(i)} (\mathbf{X} | Z^{(i)}) \delta \mathbf{X} }, \label{eqn:parallel_multi_sensor_object_bayes}
\end{align}
where
\begin{align}
\boldsymbol{\pi}^{(i)} (\mathbf{X} | Z^{(i)}) &= \frac{g^{(i)} ( Z^{(i)} | \mathbf{X} ) \boldsymbol{\pi}_{+} (\mathbf{X})}{\int g^{(i)} ( Z^{(i)} | \mathbf{X} ) \boldsymbol{\pi}_{+} (\mathbf{X}) \delta \mathbf{X} }
\end{align}
is a measurement-updated distribution produced using only the $i^{\text{th}}$ sensor.
These measurement-updated distributions can be calculated independently and in parallel, before being merged to produce the posterior distribution.
In general, this merging of measurement-updated distributions is prohibitively expensive, as it yields the exact posterior distribution.
However, if we approximate both the prior and measurement-updated distributions using LMB distributions, then merging the distributions is computationally efficient and yields an LMB posterior distribution. 

We now set about deriving the PU-LMB filter equations. We assume the prior density is the LMB distribution given by
\begin{align}
\boldsymbol{\pi}_{+} (\mathbf{X}) &= \Delta (\mathbf{X}) w_{+}(\mathcal{L} (\mathbf{X})) \prod_{(x,\ell) \in \mathbf{X} } p_{+}^{(\ell)} (x), \\
w_{+}(L) &= \prod_{j \in \mathbb{L}_{+}} \big( 1 - r_{+}^{(j)} \big)  \prod_{\ell \in L} \frac{1_{\mathbb{L}_{+}} (\ell) r_{+}^{(\ell)} }{1 - r_{+}^{(\ell)}} .
\end{align}
Under the standard multi-object measurement model, every measurement-updated distribution is an LMB mixture.
The posterior distribution given in Equation \ref{eqn:parallel_multi_sensor_object_bayes} requires each measurement-updated distribution to be divided by the prior.
If no intermediate approximations are applied to the measurement-updated distribution, then this division yields a sensor's (scaled) likelihood function.
Calculating the posterior by multiplying the prior and these likelihood functions is computationally demanding, as it requires the processing of many mixture components.
To avoid this computation, we approximate every measurement-updated LMB mixture distribution using an LMB distribution, which is given by
\begin{align}
\boldsymbol{\pi}^{(i)} (\mathbf{X} | Z^{(i)}) &= \Delta (\mathbf{X}) w^{(i)} (\mathcal{L} (\mathbf{X})) \prod_{(x, \ell) \in \mathbf{X} }  p^{(i, \ell)} (x), \\
w^{(i)} (L) &= \prod_{j \in \mathbb{L}_{+}} \big( 1 - r^{(i, j)} \big) \prod_{\ell \in L} \frac{1_{\mathbb{L}_{+}} (\ell) r^{(i, \ell)} }{1 - r^{(i, \ell)}},
\end{align}
where  $r^{(i, \ell)}$ and $p^{(i, \ell)} (x)$ are respectively an object's  measurement-updated existence probability and spatial distribution updated using sensor $i$.
In Equation \ref{eqn:parallel_multi_sensor_object_bayes}, all $S$ measurement-updated LMB distributions are then merged together to approximate the multi-sensor, multi-object Bayes posterior distribution.
As every distribution is an LMB, the approximate posterior is also an LMB distribution parameterised by
\begin{align}
\boldsymbol{\pi} =  \{ ( r^{(\ell)}, p^{(\ell)} )  \}_{\ell \in \mathbb{L}_{+}},  \label{eqn:pu_lmb_post_first_def}
\end{align}
where
\begin{align}
p^{(\ell)} (x) &= \frac{1}{\eta^{(\ell)}} ( p_{+}^{(\ell)} (x) )^{1-S} \prod_{i=1}^{S} p^{(i, \ell)} (x), \label{eqn:pu_lmb_post_spat} \\
r^{(\ell)} &= \dfrac{\eta^{(\ell)} (r_{+}^{(\ell)})^{1-S}  \prod_{i=1}^{S} r^{(i, \ell)} }{ [ (1 - r_{+}^{(\ell)})^{1-S} \prod_{i=1}^{S} (1 - r^{(i, \ell)}) ] + \eta^{(\ell)} (r_{+}^{(\ell)})^{1 - S} \prod_{i=1}^{S} r^{(i, \ell)}  }, \label{eqn:pu_lmb_post_ex} \\ 
\eta^{(\ell)} &= \int \big( p_{+}^{(\ell)} (x) \big)^{1-S} \prod_{i=1}^{S} p^{(i, \ell)} (x) d x.  \label{eqn:pu_lmb_post_norm}
\end{align}

We now derive Equations \ref{eqn:pu_lmb_post_spat} to \ref{eqn:pu_lmb_post_norm}.
By our assumptions, Equation \ref{eqn:parallel_multi_sensor_object_bayes}'s numerator is  the unnormalised LMB density
\begin{align}
\hat{\boldsymbol{\pi}} (\mathbf{X} | Z) = \Delta(\mathbf{X}) \hat{w} (\mathcal{L} (\mathbf{X})) \prod_{(x, \ell) \in \mathbf{X}} p^{(\ell)} (x), \label{eqn:pu_lmb_numerator}
\end{align}
where 
\begin{align*}
\hat{w} (L) &=  w(L)^{1 - S} \prod_{i=1}^{S} w^{(i)} (L) \\
&= \prod_{j \in \mathbb{L}_{+}} [ (1 - r_{+}^{(j)})^{1-S} \prod_{i=1}^{S} (1 - r^{(i, j)}) ] \prod_{\ell \in L} \dfrac{1_{\mathbb{L}_{+}}(\ell) \eta^{(\ell)} (r_{+}^{(\ell)})^{1 - S} \prod_{i=1}^{S} r^{(i, \ell)} }{ (1 - r_{+}^{(\ell)})^{1-S} \prod_{i=1}^{S} (1 - r^{(i, \ell)}) } \numberthis
\end{align*}
and an object's posterior spatial distribution (Equation \ref{eqn:pu_lmb_post_spat}) follows from some basic algebra.
To determine the Bayes normalisation constant, Equation \ref{eqn:parallel_multi_sensor_object_bayes}'s denominator, we require the following two results:
\begin{enumerate}
\item If $h(x)$ is a real-valued function and $X$ is a finite set, then \cite[pp. 61]{mahler_book_2014}
\begin{align}
\sum_{W \subseteq X} \prod_{w \in W} h(w) = \prod_{x \in X} (1 + h(x)), \label{eqn:power_functional_identity}
\end{align}
where $\prod_{w \in W} h(w) = 1$ if $W = \emptyset$.
\item If $w: \mathcal{F} (\mathbb{L}) \rightarrow \mathbb{R}$ and $g: \mathbb{X} \times \mathbb{L} \rightarrow \mathbb{R}$ are integrable on $\mathbb{X}$, then \cite{vo_vo_lrfs_2013}
\begin{align}
\int \Delta (\mathbf{X}) w (\mathcal{L} (\mathbf{X})) \prod_{\mathbf{x} \in \mathbf{X}} g(\mathbf{x}) \delta \mathbf{X} = \sum_{L \subseteq \mathbb{L}} w(L) \prod_{\ell \in L} \int g(x, \ell) d x , \label{eqn:set_integral_lemma}
\end{align}
where $\mathcal{F} (\mathbb{L})$ denotes the collection of all finite subsets of $\mathbb{L}$.
\end{enumerate}
Using these results, the Bayes normalisation constant of the unnormalised LMB in Equation \ref{eqn:pu_lmb_numerator} is given by
\begin{align*}
\eta &= \int  \hat{\boldsymbol{\pi}} (\mathbf{X} | Z) \delta \mathbf{X} = \sum_{L \subseteq \mathbb{L}_{+}} \hat{w} (L) \\
&= \prod_{j \in \mathbb{L}_{+}} [ (1 - r_{+}^{(j)})^{1-S} \prod_{i=1}^{S} (1 - r^{(i, j)}) ] \sum_{L \subseteq \mathbb{L}_{+}} \prod_{\ell \in L} \dfrac{ \eta^{(\ell)} (r_{+}^{(\ell)})^{1 - S} \prod_{i=1}^{S} r^{(i, \ell)} }{ (1 - r_{+}^{(\ell)})^{1-S} \prod_{i=1}^{S} (1 - r^{(i, \ell)}) } \\
&= \prod_{j \in \mathbb{L}_{+}} [ (1 - r_{+}^{(j)})^{1-S} \prod_{i=1}^{S} (1 - r^{(i, j)}) ] \prod_{\ell \in \mathbb{L}_{+}} \Big( 1 + \dfrac{\eta^{(\ell)} (r_{+}^{(\ell)})^{1 - S} \prod_{i=1}^{S} r^{(i, \ell)} }{ (1 - r_{+}^{(\ell)})^{1-S} \prod_{i=1}^{S} (1 - r^{(i, \ell)}) } \Big) \\
&= \prod_{\ell \in \mathbb{L}_{+}} \Big( \big[ (1 - r_{+}^{(\ell)})^{1-S} \prod_{i=1}^{S} (1 - r^{(i, \ell)}) \big] + \eta^{(\ell)} (r_{+}^{(\ell)})^{1 - S} \prod_{i=1}^{S} r^{(i, \ell)} \Big). \numberthis \label{eqn:pu_lmb_denominator}
\end{align*}
If we substitute Equations \ref{eqn:pu_lmb_numerator} and \ref{eqn:pu_lmb_denominator} into Equation \ref{eqn:parallel_multi_sensor_object_bayes}, we can obtain the posterior existence probability by simplifying the resulting LMB weight:
\begin{align*}
w(L) &= \frac{1}{\eta} \hat{w}(L) \\
&= \prod_{j \in \mathbb{L}_{+} - L} [ 1 + \frac{(1 - r_{+}^{(j)})^{1-S} \prod_{i=1}^{S} (1 - r^{(i, j)}) - \eta } { \eta} ] \prod_{\ell \in L} \dfrac{1_{\mathbb{L}_{+}}(\ell) \eta^{(\ell)} (r_{+}^{(\ell)})^{1 - S} \prod_{i=1}^{S} r^{(i, \ell)} }{ \eta  } \\
&= \prod_{j \in \mathbb{L}_{+} - L} ( 1 - r^{(j)} ) \prod_{\ell \in L} 1_{\mathbb{L}_{+}}(\ell) r^{(\ell)}, \numberthis
\end{align*}
where $r^{(\ell)}$ is given in Equation \ref{eqn:pu_lmb_post_ex}.

\subsection{Overview and approximations}
\label{subsection:pu_lmb_characteristics}
In this section, we first review the PU-LMB filter's computational steps before discussing its characteristics.
The major computational steps for an $S$-sensor PU-LMB filter are as follows:
\begin{enumerate}
\item \textbf{Prediction:} This step is equivalent to the single-sensor LMB filter prediction discussed in Section \ref{subsection:lmb_filter_pred}, and it also yields an LMB prior.
\item \textbf{Parallel measurement updates:} In parallel, the prior distribution is updated using measurements collected from each sensor resulting in $S$ measurement-updated distributions, each approximated as an LMB distribution.
Each measurement update is equivalent to the single-sensor LMB measurement update described in Section \ref{subsubsection:measurement_update}.
\item \textbf{Track merging:} All $S$ measurement-updated LMB distributions are combined to form the approximate posterior LMB distribution given in Equation \ref{eqn:pu_lmb_post_first_def}.
This is done by computing each object's posterior spatial distribution and existence probability using Equations \ref{eqn:pu_lmb_post_spat} and \ref{eqn:pu_lmb_post_ex} respectively.
\end{enumerate}

The main characteristic of the PU-LMB filter is that its measurement update is sensor order-independent and can be implemented in parallel.
This has an appealing, practical quality; however, the exact PU-LMB posterior density equations do suffer two major limitations, which can be relieved by approximation.
\begin{enumerate}
\item The computation of an object's posterior distribution (Equation \ref{eqn:pu_lmb_post_spat}) involves the division of probability distributions.
This may limit the types of distributions the filter can propagate.
It is possible to propagate GMs under linear Gaussian dynamics; however, even an extension to non-linear Gaussian dynamics is non-obvious. Gaussian mixture division may only be possible if an object's posterior spatial distribution can be factorised into a likelihood function multiplied by a prior distribution; however, local linearisation techniques, such as the unscented transform \cite{julier_ukf_1997}, may make such a factorisation impossible when they are applied to each component in a GM.
\item Propagating GMs may be prohibitively expensive.
If an object's prior spatial distribution is Gaussian, then its measurement-updated density $p^{(i, \ell)} (x)$ is a Gaussian mixture.
Track merging requires GM multiplication, which can be prohibitively expensive and, to mitigate this, we may wish to apply a mixture reduction approximation. Unfortunately, the required density division limits such approximations.
For example: If an object's prior is Gaussian, then we cannot approximate $p^{(i, \ell)} (x)$ using its weak marginal -- a Gaussian matching the GM's first two moments.
The weak marginal may have larger uncertainty than the prior and the resulting quotient will not be Gaussian, and this may cause numerical instabilities \cite[pp. 625]{koller_pgms_2009}.
\end{enumerate}
Despite the above characteristics, we give the PU-LMB filter an efficient, linear Gaussian implementation in Section \ref{section:simulation_results}, using Section \ref{section:efficient_approximation_lbp_lmb}'s LBP approximation. 

\section{The geometric average multi-sensor LMB (GA-LMB) filter}
\label{section:ci_lmb}
The PU-LMB filter's exact posterior density equations do not easily allow for tractable implementation, and, in practice, we must resort to approximation.
In this section, we develop an approximate, parallelisable, multi-sensor LMB filter based on geometric average (GA) fusion \cite{mahler_fusion_2000}, which alleviates some of the PU-LMB's issues. 
The multi-sensor posterior distribution is approximated as the weighted geometric average of the measurement-updated densities produced by each sensor.
As there is no density division, multi-sensor filters based on GA fusion can propagate a broader class of densities than those based on a parallel update. 

In the following subsection, we derive the GA-LMB  filter equations and discuss the filter's characteristics, focusing on a GM implementation.
\subsection{Derivation}
As before, we assume an $S$ sensor system, where each sensor obeys the standard multi-object measurement model.
We follow a similar approach to the previous section's derivation: 
First, we assume current time-step's prior is an LMB distribution.
Second, we again assume sensor $i$'s measurement-updated distribution is approximated by the LMB distribution with the parameter set
\begin{align}
\boldsymbol{\pi}^{(i)}  &= \{ ( r^{(i, \ell)}, p^{(i, \ell)} ) \}_{\ell \in \mathbb{L}_{+}} .
\end{align}
We now approximate the posterior distribution using a weighted geometric average of each measurement-updated LMB, which is the core step of GA fusion.
This geometric average produces the LMB distribution
\begin{align}
\boldsymbol{\pi} ( \mathbf{X} | Z ) \approx \dfrac{ \prod_{i = 1}^{S} ( \boldsymbol{\pi}^{(i)} (\mathbf{X} | Z^{(i)}) )^{\omega_{i}}  }{ \int \prod_{i = 1}^{S} ( \boldsymbol{\pi}^{(i)} (\mathbf{X} | Z^{(i)}) )^{\omega_{i}} \delta \mathbf{X}  }, \label{eqn:ci_bays}
\end{align}
where $\omega_{i} \in [0, 1]$ is a sensor weight, such that $\sum_{i=1}^{S} \omega_{i} = 1$.
We do not give extensive consideration to the selection of these weights in this paper.
Following the same derivation approach as Section \ref{section:pu_lmb}, the approximate LMB posterior is parameterised by
\begin{align}
\boldsymbol{\pi}  =  \{ ( r^{(\ell)}, p^{(\ell)} )  \}_{\ell \in \mathbb{L}_{+}}, 
\end{align}
where
\begin{align}
p^{(\ell)} (x) &= \frac{1}{\eta^{(\ell)}} \prod_{i=1}^{S} ( p^{(i, \ell)} (x) )^{\omega_{i}} , \label{eqn:ci_lmb_post_spatial} \\
r^{(\ell)} &= \dfrac{\eta^{(\ell)} \prod_{i=1}^{S} (r^{(i, \ell)} )^{\omega_{i}} }{ ( \prod_{i=1}^{S} ( 1 - r^{(i, \ell)} )^{\omega_{i}} ) +  \eta^{(\ell)} \prod_{i=1}^{S} ( r^{(i, \ell)} )^{\omega_{i}} }, \\
\eta^{(\ell)} &= \int \prod_{i=1}^{S} ( p^{(i, \ell)} (x) )^{\omega_{i}} d x.
\end{align}

\subsection{Overview and approximations}
\label{subsection:ci_lmb_characteristics}
In this section, we provide an overview of the GA-LMB filter and discuss its characteristics, limitations and possible approximation strategies.
The GA-LMB filter follows the same computational steps as the PU-LMB, which are prediction, parallel measurement update and track merging.
Only the track merging step differs, as it now approximates the posterior distribution using the weighted geometric average of each sensor's measurement-updated distributions.
Like the PU-LMB filter, the GA-LMB filter's two major characteristics are that its measurement update step is parallelisable and sensor order-independent.
Similarly, the filter's posterior density equations also exhibit issues and require drastic approximation:
\begin{enumerate}
\item Computing an object's posterior spatial distribution (Equation \ref{eqn:ci_lmb_post_spatial}) requires the exponentiation of densities.
In GA fusion, the following approximation is often applied to GMs \cite{li2019second, battistelli2013consensus, li2020arithmetic}: 
\begin{align}
\big( \sum_{i=1}^{n} \alpha_{i} \mathcal{N} ( x; \mu_{i}, \Sigma_{i}) \big)^{\omega} \approx \sum_{i=1}^{n} \big( \alpha_{i} \mathcal{N} ( x; \mu_{i}, \Sigma_{i}) \big)^{\omega}
\end{align}
where $\omega, \alpha_{i} \in [0, 1]$, such that $\sum_{i=1}^{n} \alpha_{i} = 1$, and $\mathcal{N} (x ; \mu, \Sigma)$ denotes a Gaussian with mean $\mu$ and covariance $\Sigma$. 
\item Like the PU-LMB filter, the propagation of GMs can still be prohibitively expensive, as an object's posterior spatial distribution is a product of GMs.
As we do not divide probability distributions, we can apply intermediate approximations to an object's measurement-updated distribution before merging the distributions together. 
\item Assuming sensor independence, the GA-LMB filter provides a poor estimate of an object's posterior covariance, even under ideal conditions.
Consider a single object which is tracked by $S$ identical, linear-Gaussian sensors, where, for each sensor, no clutter is detected and the object's measurement is always detected, i.e. $P_{D} (x, \ell) = 1$.
If the object's prior distribution is Gaussian, then its measurement-updated distribution is Gaussian, $p^{(i, \ell)} (x) = \mathcal{N} (x; \mu_{i}, \Sigma)$.
This measurement-updated distribution is equivalent to updating the object's prior using the Kalman filter update and sensor $i$'s measurement.
In this specific example, if each sensor is given equal weighting, $\omega_{i} = \frac{1}{S}$, then the object's posterior distribution is given by
\[ p ^{(\ell)} (x) = \frac{1}{\eta^{(\ell)}} \prod_{i=1}^{S} ( p^{(i, \ell)} (x) )^{\frac{1}{S}} = \mathcal{N} (x ; \frac{1}{S} \sum_{i=1}^{S} \mu_{i}, \Sigma). \]
Here the $S$ sensor GA-LMB filter is only capable of reducing the uncertainty in an object's state by a single Kalman filter update.
\end{enumerate}
Despite the above concerns, we use Section \ref{section:efficient_approximation_lbp_lmb}'s LBP approximation to give the GA-LMB filter an efficient, linear Gaussian implementation in the next section. 

\section{Implementation and experiments}
\label{section:simulation_results}
The PU- and GA-LMB filter formulations both require drastic approximation for tractable implementation.
In this section, we provide both filters with a proof-of-concept implementation, which verifies their potential using simulated data.
Both filters' relative performances and computational complexities are compared to two other LMB filters:
An LMB filter based on Vo et al.'s efficient Gibbs sampling implementation of the multi-sensor GLMB filter \cite{vo_multisensor_2017}, and an iterated-corrector (IC) LMB filter.
We refer to the first filter as the multi-sensor Gibbs sampler LMB (MGS-LMB) filter, and it represents a near-optimal implementation of a multi-sensor LMB filter.
The IC-LMB filter performs each sensor's measurement update in succession using our LBP approximation and it represents a typical implementation of a multi-sensor LMB filter.
\subsection{Implementation}
We give all four filters implemented in this section, the PU-, GA-, MGS- and IC-LMB filters, a linear Gaussian implementation.
Beyond the necessary differences, we attempt to keep the implementations as similar as possible.
We compare the filters on a $[-100~\si{\metre}, 100~\si{\metre}] \times [-100~\si{\metre}, 100~\si{\metre}]$ surveillance region using an experimental set-up that is common in the literature \cite{vo_multisensor_2017, yang_wang_lbp_2018}.
An object's state 
$x_{k} = \begin{bmatrix}
p_{x}, p_{y}, v_{x}, v_{y}
\end{bmatrix}^{T}$ 
at time $k$ describes its position and velocity in two-dimensional coordinates, and the object's stochastic motion model is described by the Markov density $f(x_{k} | x_{k-1}) = \mathcal{N} (x_{k} ; A x_{k-1}, R)$, where
\begin{align}
A = \begin{bmatrix}
1 & T  \\
0 & 1
\end{bmatrix}  \otimes I_{2 \times 2} \text{, }
R = r \begin{bmatrix}
T^{3} / 3 & T^{2} / 2 \\
T^{2} / 2 & T
\end{bmatrix}  \otimes I_{2 \times 2}, \label{eqn:motion_two}
\end{align}
$T = 1~\si{\second}$, $r=0.1$, and all objects' have a constant survival probability of $P_{S} (x, \ell) = 0.95$.
There are four spawning locations and each new object has an existence probability $r_{B}^{(\ell_{i})} = 0.03$ and a Gaussian spatial distribution $p_{B}^{(\ell_{i})} (x) = \mathcal{N} (x; \mu_{i}, \Sigma)$, where the means are $\mu_{1} = \begin{bmatrix} -80,-20, 0, 0 \end{bmatrix}^{T}$, $\mu_{2} = \begin{bmatrix} -20, 80, 0 , 0\end{bmatrix}^{T}$, $ \mu_{3} = \begin{bmatrix} 0, 0,  0,  0 \end{bmatrix}^{T} $, $ \mu_{4} = \begin{bmatrix} 40, -60, 0, 0  \end{bmatrix}^{T} $, and $ \Sigma = \text{diag} \big( \begin{bmatrix} 10, 10, 10, 10 \end{bmatrix}^{T} \big)^{2} . $
Objects with a posterior existence probability of $r^{(\ell)} < 10^{-4}$ are discarded.

In the next subsection, our simulations involve two independent sensors both of whose field-of-view (FOV) encompasses the entire surveillance region.
The first sensor's measurement model is described by $g_{1}(z | x_{k}) = \mathcal{N} (z ; C x_{k},  4 I_{2 \times 2})$, where $C = \begin{bmatrix} 1 & 0 \end{bmatrix} \otimes I_{2 \times 2}$, and each object in the sensor's FOV has a constant detection probability $P_{D, 1} (x, \ell) = 0.67$.
The second sensor's measurement model is described by $g_{2}(z | x_{k}) = \mathcal{N} (z ; C x_{k},  I_{2 \times 2})$, and each object in the sensor's FOV has a detection probability $P_{D, 2} (x, \ell) = 0.75$.
For both sensors, the clutter is generated by a Poisson process with a uniform density over the surveillance region and the expected number of clutter returns is $\lambda = 20$. 
For the GA-LMB filter, all sensors are given equal weighting in its geometric average approximation.
For the IC-LMB filter, the first sensor's measurement update is performed first.

We now discuss each filter's computational steps and their approximations.
We first discuss the PU-, GA- and IC-LMB filters' implementations, and then we discuss the MGS-LMB filter's implementation.
As discussed in Sections \ref{subsection:pu_lmb_characteristics} and \ref{subsection:ci_lmb_characteristics}, the PU- and GA-LMB filters have the following computational steps: prediction, parallel measurement update, and track merging.
The IC-LMB filter has an equivalent prediction step, but its measurement update is sequential and no track merging is required.
For the PU-, GA-, and IC-LMB filters, we approximate each sensor's measurement-updated LMB distribution using Section \ref{subsection:approximate_implemtation}'s LBP algorithm.
If a sensor collects $m$ measurements and an object's prior spatial distribution is Gaussian, then its measurement-updated spatial distribution is an $(m+1)$-component GM.
We approximate the PU-LMB filter's measurement-updated distributions using the GM component with the largest weight; this ensures the density division required in the track merging step is always valid.
For the GA- and IC-LMB filters, we approximate the measurement-updated spatial distribution using its weak marginal.
For the PU-, GA-, and IC-LMB filters, our approach ensures that an object's spatial distribution is always Gaussian.

We implemented the MGS-LMB filter using Vo et al.'s suboptimal implementation of the multi-sensor Gibbs sampler GLMB filter \cite{vo_multisensor_2017}.
This suboptimal GLMB filter achieves near-optimal results, but at significantly reduced computational cost.
At each time-step, the Gibbs sampling algorithm generates 1000 samples.
Each sample describes a valid multi-sensor data association event, and only distinct samples are kept.
The samples are used to approximate the posterior GLMB distribution, and it follows this distribution can have up to 1000 components.
In our MGS-LMB filter implementation, this posterior GLMB is then approximated as an LMB distribution, where every object's spatial distribution is a GM.
If an object's posterior spatial distribution contains more than ten components, then only the ten Gaussians with the largest weights are kept.

To evaluate the filters' tracking performances in the next subsection, we use Vo et al.'s state extraction algorithm \cite{vo_vo_2009_cbmember} to obtain a posterior state estimate.
In our implementations of the PU-, GA- and IC-LMB filters, a posterior LMB distribution is parametrised by $\boldsymbol{\pi} = \{ ( r^{(\ell)}, p^{(\ell)} ) \}_{\ell \in \mathbb{L}_{+}},$ where $ p^{(\ell)} (x) = \mathcal{N} (x; \mu_{\ell}, \Sigma_{\ell}). $
First, we determine the maximum-a-posteriori (MAP) estimate of the LMB density's cardinality distribution, then we select the $n^{\text{MAP}}$ objects with the largest existence probabilities $r^{(\ell)}$ and group their labels into the set $L$.
The posterior state estimate set is given by $\mathbf{\hat{X}} = \{ (\mu_{\ell}, \ell) \}_{\ell \in L}$, and, as we discuss in the next subsection, we also extract the set of Gaussian distributions $\hat{P} = \{ \mathcal{N} (\cdot; \mu_{\ell}, \Sigma_{\ell})  \}_{\ell \in L}.$
We apply the same state extraction algorithm to the MGS-LMB filter; however, the Gaussian component with the largest weight in each object's posterior spatial GM distribution is used to represent the object's state.

\subsection{Tracking performance}
\label{subsection:tracking_performance}
The purpose of this section's experiments is to determine whether the approximate PU- and GA-LMB filters are capable of tracking objects under ideal conditions.
All simulated scenarios are based on the same ground truth: Ten objects are tracked using two identical, independent sensors over a $100~\si{\second}$ interval using the aforementioned experimental setup.
Figures \ref{fig:pu_lmb_sim_ex_a} and \ref{fig:pu_lmb_sim_ex_b}  
\iftrue
\begin{figure}
\centering
\begin{subfigure}[b]{0.5\linewidth}
  \centering
  \includegraphics[width=5.5cm, height=5.5cm]{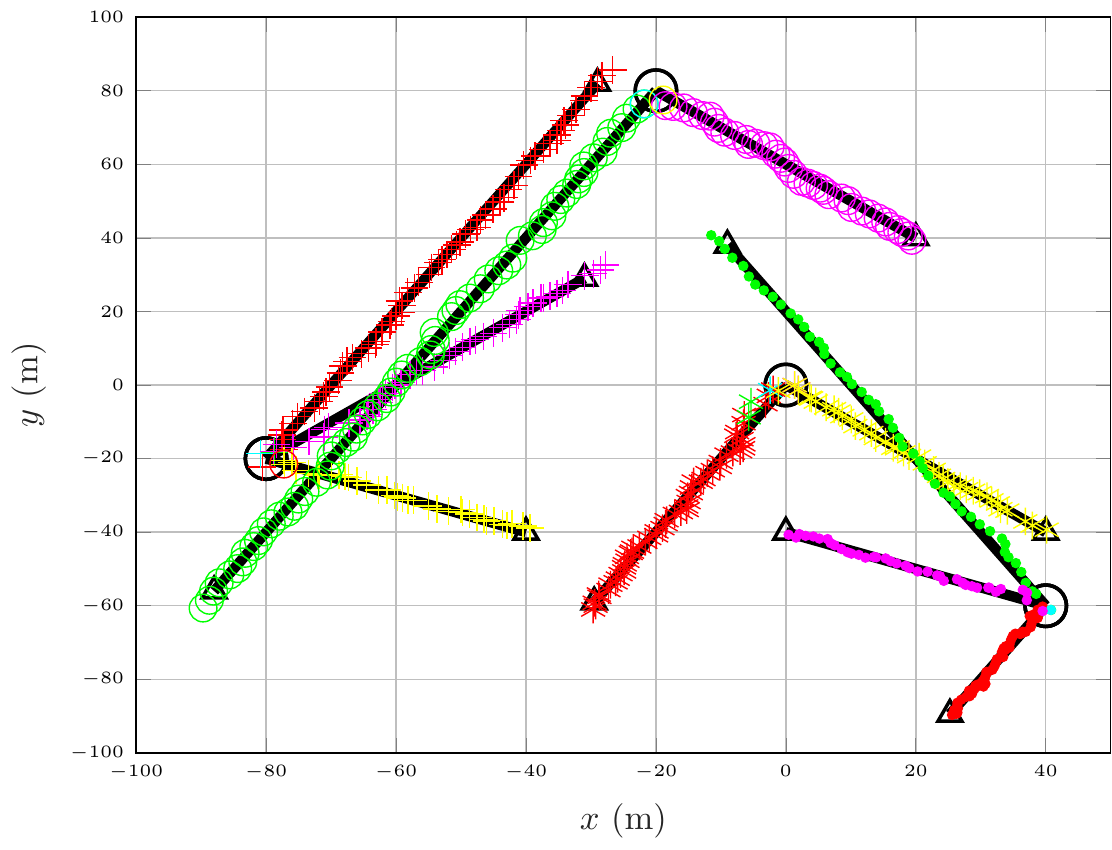}
  \caption{PU-LMB}
  \label{fig:pu_lmb_sim_ex_a}
\end{subfigure}
\begin{subfigure}[b]{0.49\linewidth}
  \centering
  \includegraphics[width=5.5cm, height=5.5cm]{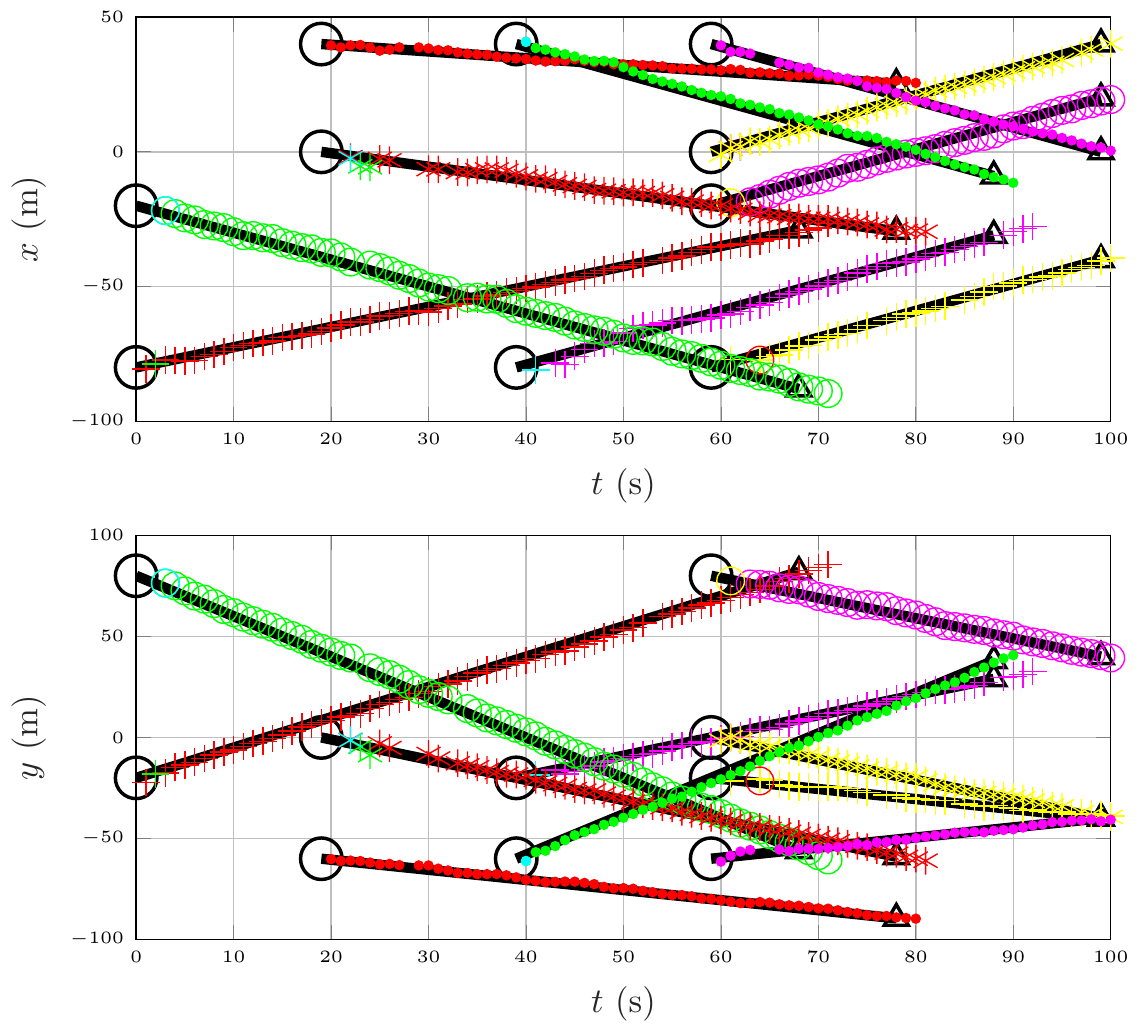}
  \caption{PU-LMB}
  \label{fig:pu_lmb_sim_ex_b}
\end{subfigure} 
\begin{subfigure}[b]{0.5\linewidth}
  \centering
  \includegraphics[width=5.5cm, height=5.5cm]{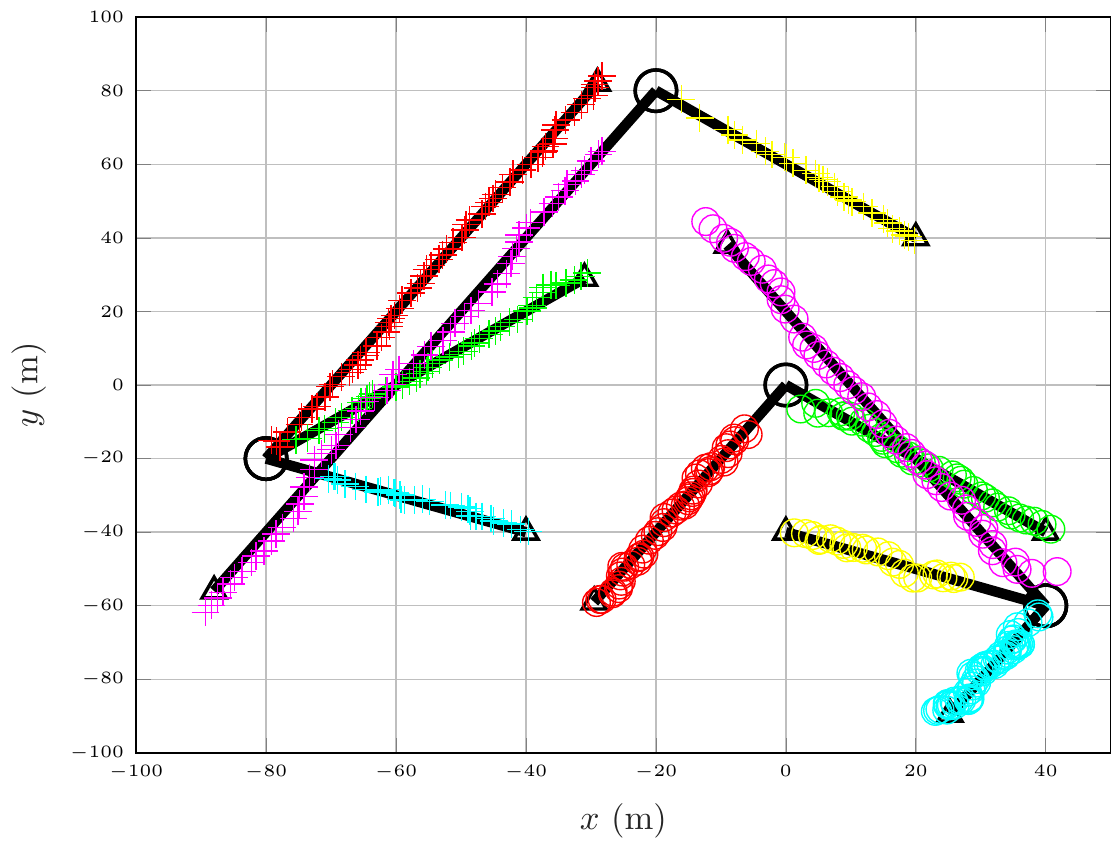}
  \caption{GA-LMB}
  \label{fig:ci_lmb_sim_ex_a}
\end{subfigure} 
\begin{subfigure}[b]{0.49\linewidth}
  \centering
  \includegraphics[width=5.5cm, height=5.5cm]{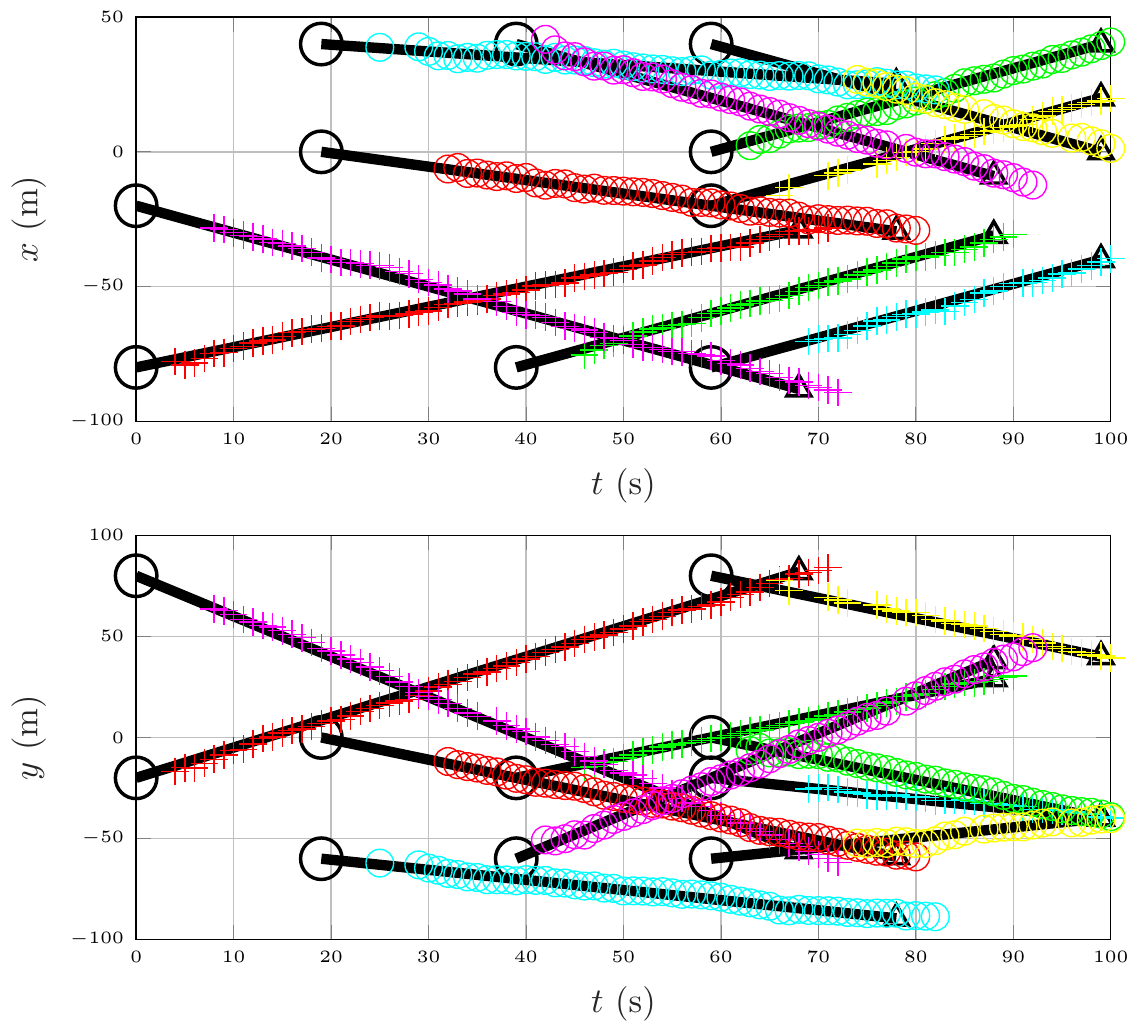}
  \caption{GA-LMB}
  \label{fig:ci_lmb_sim_ex_b}
\end{subfigure} 
\caption{The PU- and GA-LMB filter tracking ten objects using two sensors.
The start of an object's ground truth trajectory is denoted using a circle, its termination with a triangle, and its path with a solid black line.
The filters' state estimates are represented using the coloured plots, where markers with the same colour represent continuing trajectories.
Occasionally multiple tracks are initiated for each newly appearing object.
However, these false tracks quickly terminate.}
\label{fig:pu_lmb_sim_ex}
\end{figure}
\fi
depict both the objects' ground truth trajectories and the PU-LMB's state estimates.
In this example, the PU-LMB successfully tracks all objects without any track switching occurring during the multiple trajectory intersection points.
Figures \ref{fig:ci_lmb_sim_ex_a} and \ref{fig:ci_lmb_sim_ex_b} show the GA-LMB also successfully tracking the objects.
To evaluate the strength of our filters' approximations, we use two forms of the optimal subpattern assignment (OSPA) metric.
The OSPA metric is a compound metric expressing a state estimate set's ``per-object" error distance; it captures the difference in cardinality and state between two finite state estimate sets~\cite{vo_vo_ospa_2008}.
Figure \ref{fig:high_detection_prob_ospa} 
\iftrue
\begin{figure}[tbp]
	\centering
	\includegraphics[scale=0.8]{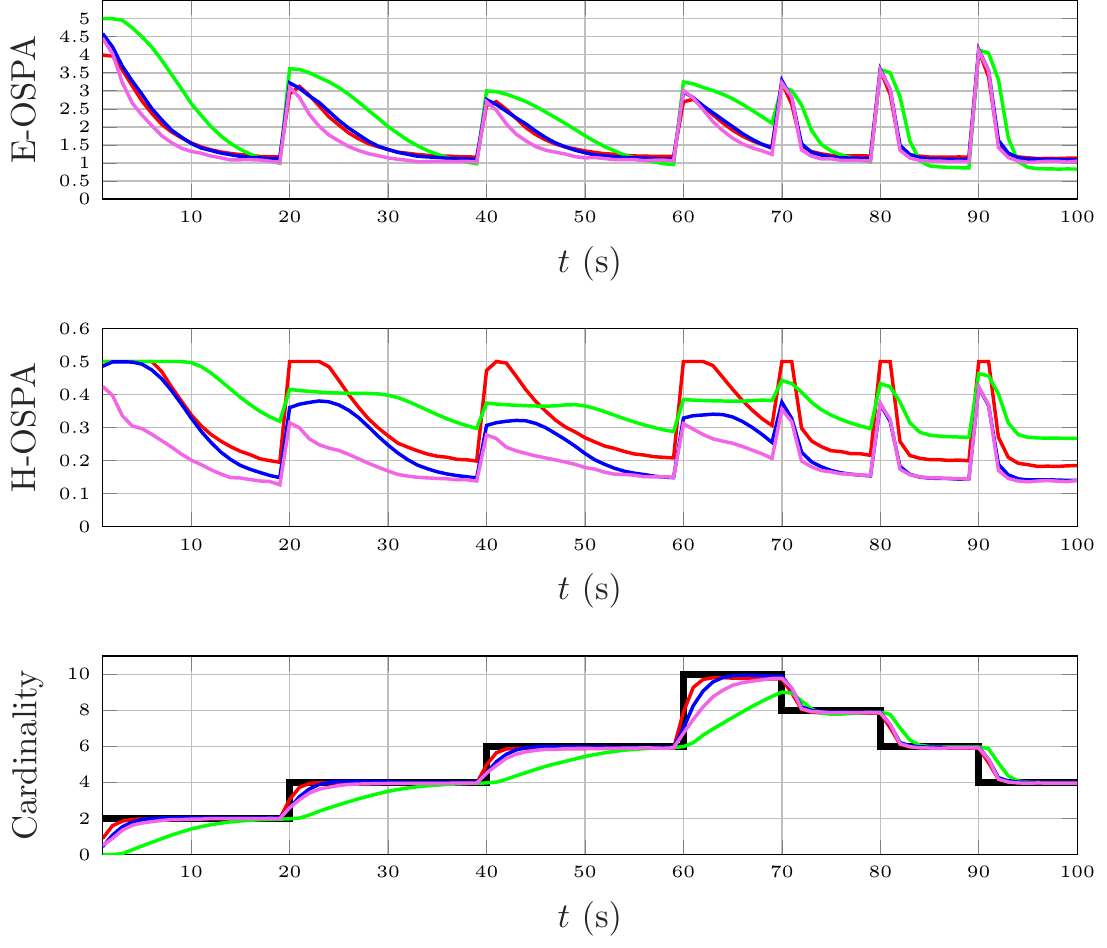}
	\caption{The PU-, GA-, IC-, and MGS-LMB filters' performance results, as compared to the ground truth, and using an average of $5000$ simulations.
	The PU-LMB filter's results are given by the red line, the GA-LMB filter by the green line, the IC-LMB filter by the blue line, and the MGS-LMB filter by the purple line.
	In the cardinality plot, the ground truth is represented by the black line. The E-OSPA uses the parameters $c_{E} = 5~ \si{\metre}$ and $p_{E} = 2$, while H-OSPA uses the parameters $c_{H} = 0.5$ and $p_{H} = 2$.}
	\label{fig:high_detection_prob_ospa}
\end{figure}
\fi
shows the average Euclidean and Hellinger OSPA distances (E-OSPA and H-OSPA respectively) for the given ground truth, where the PU-LMB and GA-LMB are both compared to the MGS- and IC-LMB filters.
The OSPA represents both a filter's error in object number and localisation using single value, where a localisation error is an error in an object's kinematic state.
To distinguish between these two type of errors, we have included a cardinality plot in the OSPA figure.
Both OSPA metrics spike for all the filters when a new object appears, indicating a likely cardinality error, as the filters do not yet represent the new object in their state estimate sets.
If the filters correctly estimate the cardinality, then the OSPA metric only represents localisation error, providing the average per-object error.
In this case, if the E-OSPA is $2~ \si{\metre}$, then this indicates the average Euclidean distance between each object's localisation estimate and its corresponding ground truth state is $2~ \si{\metre}$.
The E-OSPA's localisation has a cut-off parameter that indicates the largest error we deem acceptable, in this case $5~ \si{\metre}$.
Similarly, the H-OSPA's localistion component measures the average Hellinger distance between each object's Gaussian state estimate and its corresponding optimal state estimate \cite{clark_hellinger_ospa_2011}.
Here, an optimal state estimate is a Gaussian distribution produced by a multi-sensor Kalman filter with known data association.
The Hellinger distance measures the ``similarity" between two distributions and it has a maximum value of $1$; indicating total dissimilarity between distributions, we chose an H-OSPA cut-off of $0.5$.
As can be seen from the E-OSPA results, the filters all provide relatively accurate position estimates.
From the H-OSPA results, it can be seen the GA-LMB filter provides a poor covariance estimate, as expected.
The GA-LMB filter also provides the worst cardinality estimate of all the filters.
However, once the GA-LMB filter provides the correct cardinality estimate, then its average E-OPSA metric is lower than the other filters' metrics.
This implies the GA-LMB does track all the objects present; however, they are not always accurately represented in the filter's extracted posterior state estimate set.
The PU-LMB filter provides a nearly equivalent state estimate to the IC-LMB filter; however, the PU-LMB's initial estimates for a newly appearing object are consistently less accurate.
The MGS-LMB filter provides the most accurate state estimates, as it is the least approximate.

The experiments indicate that the GA-LMB's performance is worse than the other filters' performances.
However, the GA-LMB filter is more readily applicable to non-linear dynamics than the PU-LMB filter, as its formulation does not require the division of probability densities.
\subsection{Computational complexity}
A single-sensor LMB filter implemented using Section \ref{subsection:approximate_implemtation}'s LBP algorithm takes $O (k n m)$ time \cite{Reuter_gibbs_2017}, where $k$, $n$ and $m$ are the respective numbers of LBP iterations, objects and measurements at a given time-step.
For an $S$-sensor system, where each sensor collects $m_{i}$ measurements and $m_{\text{max}}$ is the largest number of measurements collected, it follows that the  PU- and GA-LMB filters take $O(k n m_{\text{max}})$ time, if each sensor's measurement update is computed in parallel.
In contrast, the IC-LMB takes $O(k n \prod_{i=1}^{S} m_{i})$ time, and the MGS-LMB filter takes $O(p n^{2} \prod_{i=1}^{S} m_{i})$ time, where $p$ is the number of samples.
In this subsection, we verify the filters' asymptotic computational complexity by varying the number of sensors in a given simulation.

We use the same experimental setup as before; however, we now track 50 ever-present objects on a $[-500~\si{\metre}, 500~\si{\metre}] \times [-500~\si{\metre}, 500~\si{\metre}]$ observation space, where the spawning locations are chosen randomly over this region.
We have increased the simulation length to $3000$ time-steps and we have lowered the update rate ($T$ in Equation \ref{eqn:motion_two}) to $20~\si{\milli \second}$.
We have also increased the detection probability, and it is now given by $P_{D} (x, \ell) = 1$ for each sensor.
The experiments were implemented in the Julia v$1.05$ language, and run on a personal computer with an Intel\textsuperscript{\textregistered} Core i$7$-$6700k$ CPU running at $4~\si{\giga \hertz}$, with $4$ cores and $8$ logical processors available.
The $3000$ time-step run-times of $1000$ simulations were averaged to produce a single data point.
As Figure \ref{fig:sensor_number}
\iftrue
\begin{figure}
\centering
\begin{subfigure}[b]{0.5\linewidth}
  \centering
  \includegraphics[width=0.9\linewidth]{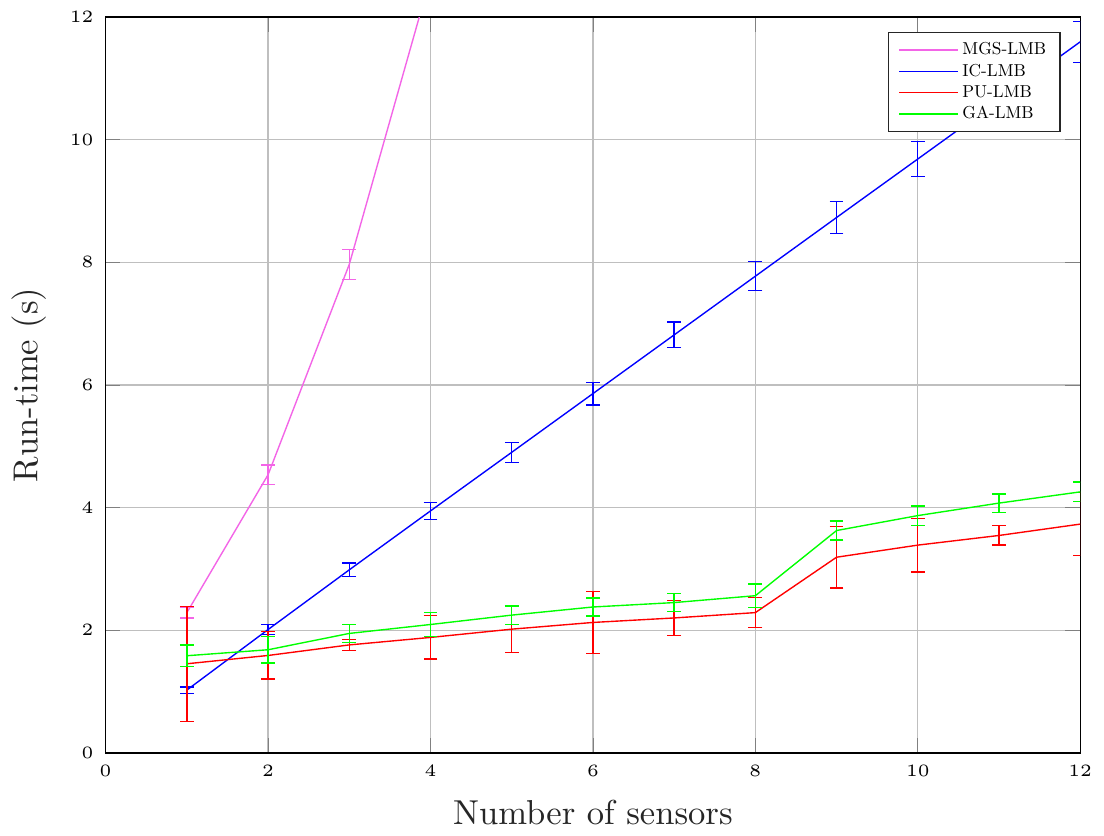}
  \caption{}
  \label{fig:sensor_number_zoomed_in}
\end{subfigure}%
\begin{subfigure}[b]{0.5\linewidth}
  \centering
  \includegraphics[width=0.9\linewidth]{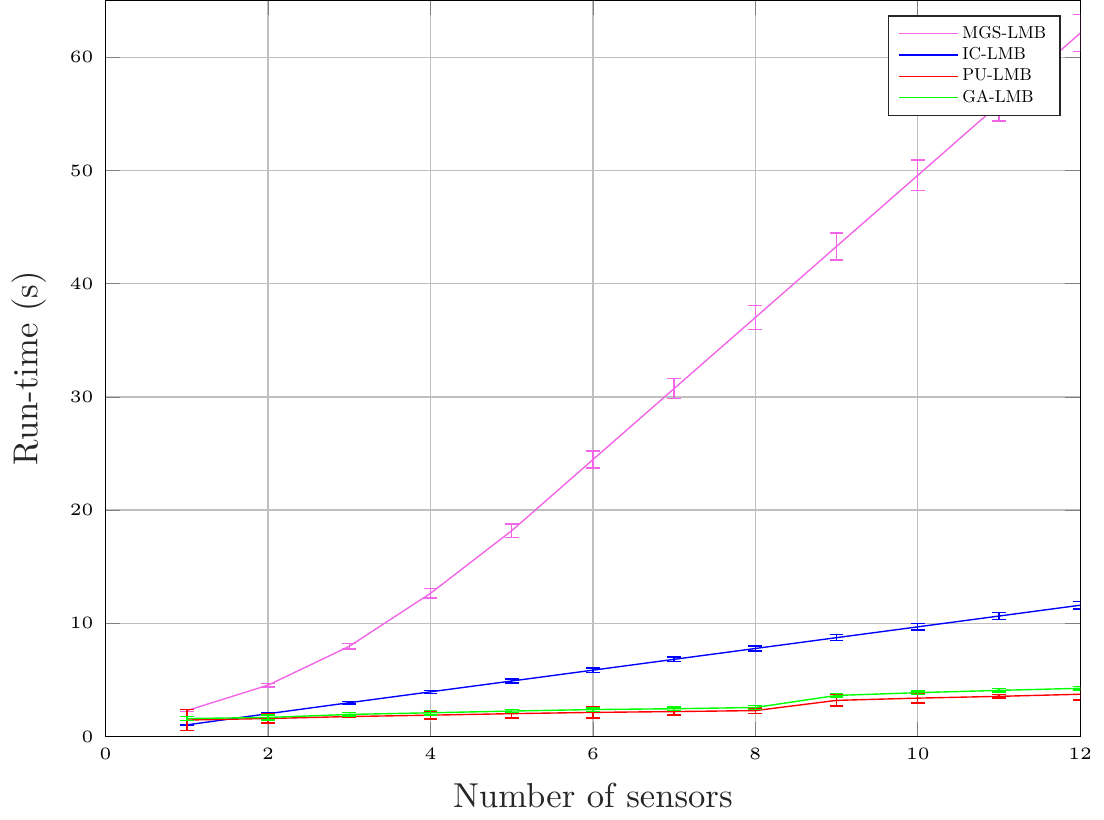}
  \caption{}
  \label{fig:sensor_number_zoomed_out}
\end{subfigure} 
	\caption{The run-time of the filters plotted against sensor number.
	A total of eight logical threads are available, and a maximum of eight measurement-updates can be completed simultaneously in parallel.
	This results in a doubling of the run-time for the parallel PU- and GA-LMB filters when nine or more sensors are simulated.}
\label{fig:sensor_number}
\end{figure}
\fi
indicates, the MGS- and IC-LMB filters' run-times are linear in sensor number, while the PU- and GA-LMB filters' are nearly constant.
Due to practical limitations, the PU- and GA-LMB filters both exhibit a run-time increase when the maximum number of available threads is exceeded.

Based on the simulations, the PU- and GA-LMB filters exhibit a low computational cost.
This makes them suitable for tracking a large number of objects using many sensors in real-time.
There are situations when filter execution time is as important as filter accuracy.
We now discuss an application example from the sports entertainment industry, where golf ball trajectories at a driving range are tracked and reported back to a player at a particular bay \cite{inrangeGolf}.
A large driving range may have up to 130 bays and could be surveilled by up to 14 sensors.
As driving bay location is tied to object identity, this translates to a maximum of 130 newly appearing objects at each time-step.
A player expects to see their shot reported back to them the moment it strikes the ground.
Waiting for a shot report, regardless of how accurately it may have been tracked, has a negative impact on player enjoyment.

\section{Conclusion}
This paper proposed and implemented two efficient, parallelisable approximations of the multi-sensor LMB filter, the PU- and GA-LMB filters.
The PU-LMB filter formulation resulted from the direct manipulation of the multi-sensor, multi-object Bayes filter posterior's distribution.
It approximated each sensor's measurement-updated distribution using an LMB distribution, resulting in an LMB posterior distribution.
Unfortunately, the PU-LMB filter requires the division of probability distributions and its extension beyond linear-Gaussian mixture implementations is non-obvious.
The GA-LMB filter formulation approximated the multi-sensor, multi-object Bayes filter's posterior distribution by first approximating each sensor's measurement-updated distribution as an LMB density, and then approximating the posterior distribution as the weighted geometric average of these measurement-updated distributions.
Like the PU-LMB filter, this formulation also resulted in an approximate LMB posterior density. However, the GA-LMB filter can be used under non-linear conditions.
In addition to further approximations, both proposed LMB filters made use of the same LBP message passing algorithm to approximate their posterior densities.
Unlike Williams et al.'s original LBP data association algorithm, we derived the algorithm using a cluster graph and binary association variables.
Our approach allows both filters to have a constant complexity in the number of sensors, and linear complexity in both number of objects and measurements.
This is an improvement on both the MGS- and IC-LMB filters, which both have linear complexities in the number of sensors.
The experiments reveal that the accuracy of the PU- and GA-LMB filters is not significantly worse than the IC-LMB filter, and both filters are less accurate than the MGS-LMB filter.
However, the PU- and GA-LMB filters have a constant computational complexity in the number of sensors, rather than linear.
On simulated data, the PU-LMB filter is more accurate than the GA-LMB filter, providing a more accurate covariance and cardinality estimate.
Both filters are of interest in scenarios where many objects are tracked using several sensors, and filter run-time is more important than filter accuracy.

\section*{Acknowledgment}
This work was supported by \href{https://www.inrangegolf.com/}{Inrange Golf}.
We also thank the anonymous reviewers for the suggestions which helped clarify many points.

\bibliography{mybibfile}
\end{document}